\documentclass[journal]{IEEEtran}
\usepackage[utf8]{inputenc}
\usepackage{amsmath}
\usepackage{algorithmic}
\usepackage{array}
\usepackage{comment}
\ifCLASSOPTIONcaptionsoff
\usepackage[nomarkers]{endfloat}
\let\MYoriglatexcaption\caption
\renewcommand{\caption}[2][\relax]{\MYoriglatexcaption[#2]{#2}}
\fi
\usepackage{url}
\usepackage{cite}
\usepackage{url}
\usepackage{algorithm}
\usepackage{algorithmic}
\usepackage{acronym}
\usepackage[table]{xcolor}
\usepackage{tablefootnote}
\usepackage{siunitx}

\usepackage{smartdiagram}
\usepackage{metalogo}
\usepackage{tikz}
\usetikzlibrary{mindmap,trees}
\usepackage{tikz}
\usetikzlibrary{arrows,shapes,positioning,shadows,trees}

\tikzset{
  basic/.style  = {draw, text width=3cm, drop shadow, font=\sffamily, rectangle},
  root/.style   = {basic, rounded corners=1pt, thin, align=center,
                   fill=green!30},
  level 2/.style = {basic, rounded corners=8pt, thin,align=center, fill=green!60,
                   text width=6em},
  level 3/.style = {basic, thin, align=left, fill=pink!60, text width=7em}
}

\acrodef{adc}[ADC]{Analog-to-Digital Converter}
\acrodef{tmd}[TMD]{Transport Mode Detection}
\acrodef{keh}[KEH]{Kinetic Energy Harvesting}
\acrodef{seh}[SEH]{Solar Energy Harvesting}
\acrodef{teh}[TEH]{Thermal Energy Harvesting}
\acrodef{rfeh}[RFEH]{RF Energy Harvesting}
\acrodef{iot}[IoT]{Internet of Things}
\acrodef{rf}[RF]{Random Forest}
\acrodef{dt}[DT]{Decision Tree}
\acrodef{svm}[SVM]{Support Vector Machine}
\acrodef{knn}[KNN]{K-Nearest Neighbor}
\acrodef{nb}[NB]{Naive Bayes}
\acrodef{apr}[APR]{Acquisition Power Ratio}
\acrodef{leds}[LEDs]{Light Emitting Diodes}
\acrodef{cv}[CV]{Cross Validation}
\acrodef{rfe}[RFE]{Recursive Feature Elimination}
\acrodef{smote}[SMOTE]{Synthetic Minority Over-sampling Technique}
\acrodef{mpp}[MPP]{Maximum Power Point}
\acrodef{mosfet}[MOSFET]{Metal Oxide Semiconductor Field Effect Transistor}
\acrodef{pmu}[PMU]{Power Management Unit}
\acrodef{emu}[EMU]{Energy Management Unit}
\acrodef{eno}[ENO]{Energy Neutral Operation}
\acrodef{hvac}[HVAC]{Heating, Ventilation, and Air Conditioning}
\acrodef{mems}[MEMS]{Micro-Electro-Mechanical Systems}
\acrodef{cps}[CPS]{Cyber-Physical System}
\acrodef{em}[EM]{Electromagnetic}
\acrodef{ewma}[EWMA]{Exponentially Weighted Moving Average}
\acrodef{ehiot}[EH-IoT]{Energy Harvesting based IoT}
\acrodef{dvfs}[DVFS]{Dynamic Voltage and Frequency Scaling}
\acrodef{tdma}[TDMA]{Time Division Multiple Access}
\acrodef{soc}[SoC]{State of Charge}
\acrodef{edf}[EDF]{Earliest Deadline First}
\acrodef{esu}[ESU]{Energy Storage Unit}
\acrodef{teg}[TEG]{Thermoelectric Power Generator}
\acrodef{wsns}[WSNs]{Wireless Sensor Networks}
\acrodef{qos}[QoS]{Quality-of-Service}
\acrodef{mdp}[MDP]{Markov Decision Process}
\acrodef{asap}[ASAP]{As Soon As Possible}
\acrodef{alap}[ALAP]{As Late As Possible}
\acrodef{edf}[EDF]{Earliest Deadline First}
\acrodef{rfid}[RFID]{Radio Frequency Identification}
\acrodef{peh}[PEH]{Piezoelectric Energy Harvester}
\acrodef{eeh}[EEH]{Electromagnetic Energy Harvester}
\acrodef{src}[SRC]{Sparse Representation based Classification}
\acrodef{adl}[ADL]{Activities of Daily Living}
\acrodef{mac}[MAC]{Medium Access Control}

\usepackage{lscape} 
\usepackage{longtable}
\usepackage{booktabs} 
\usepackage{multirow} 
\usepackage{amssymb}

\usepackage{amsmath,amssymb,amsfonts}
\usepackage{algorithmic}
\usepackage{graphicx}
\usepackage{textcomp}
\usepackage{xcolor}

\usepackage[colorlinks=false]{hyperref}

\usepackage{tabulary}
\newcolumntype{K}[1]{>{\centering\arraybackslash}m{#1}}
\usepackage{multicol}
\usepackage{url}
\title{Task Scheduling for Simultaneous IoT Sensing and Energy Harvesting: A Survey and Critical Analysis}
 \author{\IEEEauthorblockN{Muhammad Moid Sandhu~\href{https://orcid.org/0000-0001-6444-0900}{\includegraphics[scale=0.08]{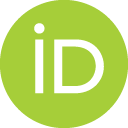}},~\IEEEmembership{Student~Member,~IEEE,} Sara Khalifa~\href{https://orcid.org/0000-0002-3417-2834}{\includegraphics[scale=0.08]{orcid.png}},~\IEEEmembership{Member,~IEEE,} \\ Raja Jurdak~\href{https://orcid.org/0000-0001-7517-0782}{\includegraphics[scale=0.08]{orcid.png}},~\IEEEmembership{Senior~Member,~IEEE,} Marius Portmann~\href{https://orcid.org/0000-0003-1852-3961}{\includegraphics[scale=0.08]{orcid.png}},~\IEEEmembership{Senior~Member,~IEEE}}\\
 
 \thanks{M. M. Sandhu is with the University of Queensland, and Commonwealth Scientific and Industrial Research Organization (CSIRO)'s Data61, Australia. \textit{Email: m.sandhu@uqconnect.edu.au}}
  \thanks{S. Khalifa is with the CSIRO's Data61, University of New South Wales, and University of Queensland, Australia. \textit{Email: sara.khalifa@data61.csiro.au}}
   \thanks{R. Jurdak is with the Queensland University of Technology and CSIRO's Data61, Australia. \textit{Email: r.jurdak@qut.edu.au}}
    \thanks{M. Portmann is with the University of Queensland, Australia. \textit{Email: marius@itee.uq.edu.au}}}
\date{}

\begin{document}

\maketitle

\begin{abstract}
The \ac{iot} has important applications in our daily lives including health and fitness tracking, environmental monitoring and transportation. However, sensor nodes in \ac{iot} suffer from the limited lifetime of batteries resulting from their finite energy availability. A promising solution is to harvest energy from environmental sources, such as solar, kinetic, thermal and radio frequency, for perpetual and continuous operation of \ac{iot} sensor nodes. In addition to energy generation, recently energy harvesters have been used for context detection, eliminating the need for additional activity sensors (e.g. accelerometers), saving space, cost, and energy consumption. Using energy  harvesters for simultaneous sensing and energy harvesting enables \textit{energy positive sensing} -- an important and emerging class of sensors, which harvest higher energy than required for signal acquisition and the additional energy can be used to power other components of the system. Although simultaneous sensing and energy harvesting is an important step forward towards autonomous self-powered sensor nodes, the energy and information availability can be still intermittent, unpredictable and temporally misaligned with various computational tasks on the sensor node. This paper provides a comprehensive survey on task scheduling algorithms for the emerging class of energy harvesting-based sensors (i.e., energy positive sensors) to achieve the sustainable operation of \ac{iot}. We discuss inherent differences between conventional sensing and energy positive sensing and provide an extensive critical analysis for devising new task scheduling algorithms incorporating this new class of sensors. Finally, we outline future research directions towards the implementation of autonomous and self-powered \ac{iot}.
\end{abstract}
\IEEEpeerreviewmaketitle

\section{Introduction}
\IEEEPARstart{W}{ith} the advancements in \ac{mems}, low power miniaturized sensors in \ac{iot} are becoming popular for monitoring the physical attributes in various applications including surveillance, smart cities, healthcare, exploration of mines, battle field monitoring and even deep sea exploration\cite{singh2014survey,lee2015internet,chen2014vision,islam2015internet,farahani2018towards,li2017iot,kim2019wearable,hassan2018kinetic,wijesundara2016design,zhang2004hardware,al2015internet}. 
However, conventional sensor nodes employ a rechargeable battery, which has limited energy storage capacity~\cite{zhang2013distributed} and thus hinders the perpetual operation of sensor nodes in \ac{iot}. 
In order to solve this problem, recently, kinetic, solar, thermal, and RF energy harvesters~\cite{penella2007review,chalasani2008survey,hinchet2015wearable} have been used to convert the environmental energy into electrical energy, to extend the battery lifetime of sensor nodes~\cite{yang2017energy,sigrist2017measurement}.
This relieves the issue of limited lifetime of batteries, and thus allows the autonomous operation of sensor nodes in \ac{ehiot} without human intervention.
\begin{figure}[t!]
\centering
\includegraphics[width=9cm, height=8cm]{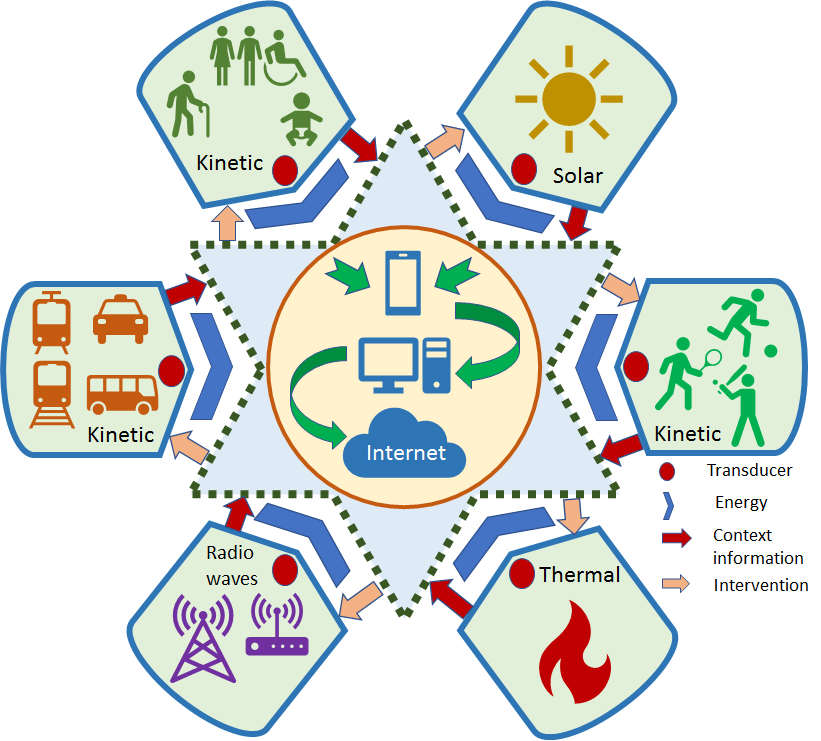}
\caption{Energy harvesters can be used as context sensors as well as source of energy in \ac{iot}}
\label{fig:intro_figure}
\vspace{-0.5cm}
\end{figure}

\begin{table*}[t!]
\caption{Comparison of this work with previous survey papers related to \ac{iot} and energy focused research}
\label{table_previous_surveys}
\centering
\begin{tabular}{llcccc}
\toprule
Year & Reference & Energy harvesting & Sensing using energy harvesting & Simultaneous sensing and energy harvesting & Task scheduling \\
\midrule\midrule
2006 & \cite{wang2006survey} & $\times$ & $\times$ & $\times$ & \checkmark \\
2008 & \cite{yu2008survey} & $\times$ & $\times$ & $\times$ & \checkmark \\
2010 & \cite{sudevalayam2010energy} & \checkmark & $\times$ & $\times$ & $\times$ \\
2013 & \cite{carrano2013survey} & $\times$ & $\times$ & $\times$ & \checkmark \\
2014 & \cite{valera2014survey} & \checkmark & $\times$ & $\times$ & $\times$ \\
2016 & \cite{bambagini2016energy} & $\times$ & $\times$ & $\times$ & \checkmark \\
2016 & \cite{shaikh2016energy} & \checkmark & $\times$ & $\times$ & $\times$ \\
2018 & \cite{adu2018energy} & \checkmark & $\times$ & $\times$ & $\times$ \\
2019 & \cite{ma2019optimizing} & \checkmark & \checkmark & $\times$ & $\times$ \\
2020 & Proposed & \checkmark & \checkmark & \checkmark & \checkmark \\
\bottomrule
\end{tabular}
\end{table*}

Recently, energy harvesters are also being used as sensors for activity detection~\cite{khalifa2015energy,lan2018entrans,ma2018solargest,khalifa2018harke}, as depicted in Fig.~\ref{fig:intro_figure}, to save sensor-related energy consumption~\cite{khalifa2018harke} that would otherwise be used for powering conventional activity sensors, such as accelerometers. Energy harvesters can be employed as energy efficient sensors to detect the underlying activity as well as a source of energy to power sensor nodes~\cite{sandhu2020ssehkeh}.
In simultaneous sensing and energy harvesting paradigm, the harvested energy can be used to at least acquire the energy harvesting signal (for context detection) leading towards energy positive sensing~\cite{sandhu2020ssehkeh} compared to conventional energy negative sensing.
This allows full utilization of energy harvesting capabilities in practical and real-world environments, leading towards autonomous operation of sensor nodes in \ac{ehiot}.


\subsection{Background} Despite the emerging importance of energy harvesting, the amount of generated energy from the environment is still insufficient to enable the \ac{eno}~\cite{baghaee2014demonstration} of miniaturized sensor nodes~\cite{sandhu2020optimal}, especially for wearable sensing devices with small form factor. There are various mechanisms that can be employed to enable the \ac{eno} of miniaturized sensor nodes such as: (i) maximizing the harvested energy by implementing optimal energy harvesting mechanisms ~\cite{sandhu2020optimal,omairi2017power} or considering multi-source energy harvesters~\cite{bandyopadhyay2012platform}, and (ii) minimizing the energy consumption~\cite{jang2017circuit} by using novel low power sensing mechanisms such as energy harvesting-based sensing~\cite{khalifa2018harke}. In addition to maximum energy harvesting and low power sensing , energy management algorithms can be employed to manage the precious harvested energy, ensuring \ac{eno} of sensor nodes~\cite{ejaz2017efficient}.
There are different types of energy management algorithms, including transmission power control~\cite{wu2017optimal,lin2016atpc,xiao2009transmission}, \ac{mac} scheduling~\cite{sherazi2018comprehensive,chang2008energy,jang2013asynchronous} and task scheduling~\cite{caruso2018dynamic,rault2014energy,xiong2011multiple}, to name a few.
\begin{figure}[t!]
\centering
\includegraphics[width=7cm, height=4cm]{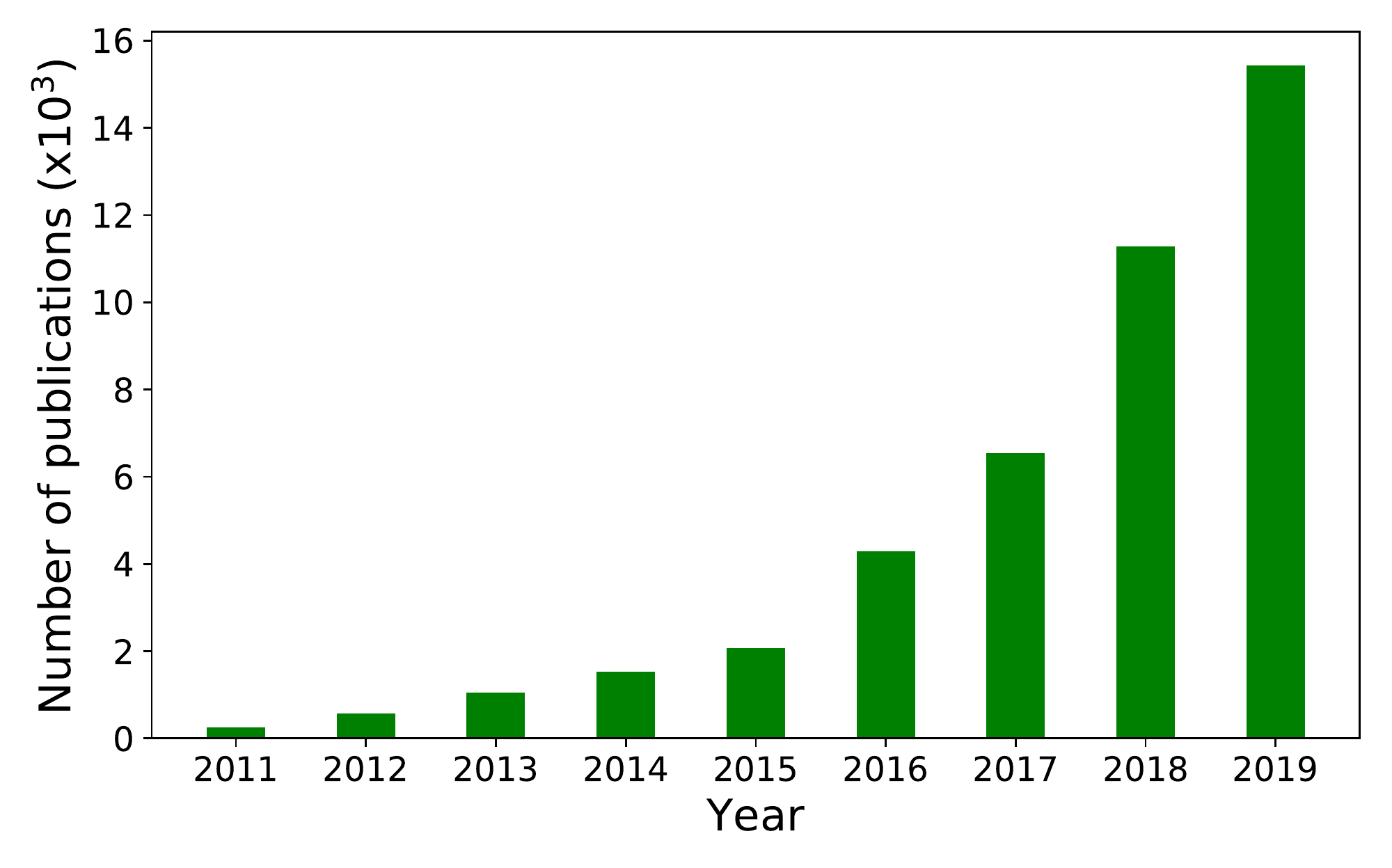}
\caption{Research trend in the previous years depicts the growing potential of publications containing the following keywords: \textit{Energy harvesting \ac{iot}, Task scheduling in \ac{iot}, and Batteryless \ac{iot}}}
\label{fig_research_papers}
\vspace{-0.5cm}
\end{figure}

Task scheduling algorithms schedule the broader set of tasks (such as sampling, processing, transmission, etc.) on the sensor node, according to the available energy budget, to prolong its operational lifetime. It is an effective method to minimize the energy consumption on the sensor node, due to its direct interaction with the \ac{esu} (i.e., battery/capacitor) and energy consumption (in executing the tasks) on the processor. 
Task scheduling algorithms are more effective to minimize the energy consumption, than other communication-focused energy management schemes (i.e., transmission power control) due to their interaction with a broader set of tasks including digitizing, sampling, and processing as well as communication. The objective of task scheduling algorithms is to run sensor nodes using the intermittent harvested energy~\cite{lucia2017intermittent,hester2017timely} to ensure \ac{eno} and achieve maximum sensing performance in \ac{iot} applications. Therefore, in this survey, we comprehensively analyse task scheduling algorithms for energy harvesting based sensing to enable the \ac{eno} of sensor nodes in \ac{ehiot}.

\begin{table}[t!]
\caption{Nomenclature}
\label{table_Nomenclature}
\centering
\begin{tabular}{ll}
\toprule
Term & Detail \\
\midrule\midrule
$V_{AC}$ & AC voltage \\
$V_{oc}$ & Open circuit voltage \\
$V_{rec}$ & Rectified voltage \\
$V_{R}$ & Resistor voltage \\
$V_{cap}$ & Capacitor voltage \\
$V_{pmu}$ & \ac{pmu} voltage \\
$P_{har}$ & Harvested power \\
$P_{acq}$ & Signal acquisition power \\
\bottomrule
\end{tabular}
\end{table}

\subsection{Motivation}
In recent years, there has been a growing trend in energy harvesting mechanisms to power \ac{iot} sensors and related task scheduling-based energy management schemes, as depicted in Fig.~\ref{fig_research_papers}\footnote{The numbers are obtained from \textit{Dimensions}\\Source: \url{https://app.dimensions.ai/discover/publication}\\Accessed on: March 16, 2020}.
Previously, energy harvesters have been used for activity detection~\cite{khalifa2018harke,lan2018entrans}, as depicted in Fig.~\ref{fig:intro_figure}, to replace conventional activity sensors, which operate using the supplied power. Some of the previous works also use multi-source energy harvesters~\cite{umetsu2019ehaas} to harvest higher energy as well as rich context information. However, only one proposal~\cite{sandhu2020ssehkeh} employs the harvested energy to power a system load. In order to make best use of energy harvesters, they must be employed as a simultaneous source of energy~\cite{ku2016advances,raghunathan2005design, kansal2003environmental, corke2007long} and context information~\cite{khalifa2018harke}. This enables energy positive sensing which harvest enough energy to acquire the energy harvesting signal, as discussed in Section~\ref{energy_positive_sensing}. Finally, task scheduling schemes can be incorporated to ensure \ac{eno} of sensor nodes using the limited and unreliable harvested energy. The ability to extract information from energy harvesting signals, and in some cases to gain energy in the process, has the potential to dramatically change the task scheduling landscape for \ac{ehiot}. Scheduling algorithms have to consider the information and energy gain (for energy positive sensing), rather than energy loss, of different sensors to achieve an objective function. This potential warrants a comprehensive survey of task scheduling algorithms to analyse their support for such decisions. 

\subsection{State-of-the-art}
There are several works in the literature that explore task scheduling in \ac{ehiot}.
Table~\ref{table_previous_surveys} compares our paper to previous survey papers related to \ac{iot} with energy focused research. Some of the previous works~\cite{wang2006survey, yu2008survey, carrano2013survey, bambagini2016energy} present extensive surveys on task scheduling schemes to minimize energy consumption in \ac{iot}. However, none of theses surveys considered energy harvesting mechanisms to power \ac{iot} sensor nodes with the associated opportunities and challenges compared to battery-powered \ac{iot}. 
Others survey papers~\cite{sudevalayam2010energy, shaikh2016energy, adu2018energy,valera2014survey} covered energy harvesting mechanisms to power \ac{iot} sensor nodes in order to enhance their operational lifetime and to reduce maintenance cost. Nevertheless, they considered conventional sensors instead of energy harvesting-based sensing. A recent comprehensive survey ~\cite{ma2019optimizing} is the first to cover energy harvesting-based sensing while optimizing sensing, computing and communication for \ac{ehiot}. However, simultaneous sensing and energy harvesting as well as the resulted energy positive sensing concept has not been discussed. Moreover, none of previous survey papers~\cite{sudevalayam2010energy, shaikh2016energy, adu2018energy,ma2019optimizing}, which focus on energy harvesting research, considered task scheduling as a crucial mechanism to manage execution of tasks under the limited and time-varying harvested energy. 

\subsection{Contributions} To address the aforementioned gaps in the literature, this work critically surveys task scheduling schemes to minimize the energy consumption for perpetual operation of sensor nodes in \ac{ehiot}, and analyses their potential to support energy-harvesting based sensors. Our contributions are as follows:
\begin{itemize}
    \item We discuss the implementation of energy harvesters as sensors and energy sources simultaneously, which is more useful in practical environments leading towards self-powered batteryless \ac{iot}. We also explore the concept of energy positive sensing, which uses the harvested energy to acquire the energy harvesting signal for sensing, eliminating the need for additional power consuming sensors.
    \item We present an extensive discussion of task scheduling based energy management algorithms for running the tasks on the resource constrained sensor node under limited and varying harvested energy due to unreliable environmental energy (i.e., kinetic, solar, thermal, RF, etc).
    \item Based on an extensive study of the literature, we comprehensively describe the key challenges and potential solutions when integrating energy positive sensing with conventional task scheduling algorithms.
    \item Finally, we present the future research directions to enable the sustainable and autonomous operation of batteryless sensor nodes in \ac{ehiot}.
\end{itemize}

The remainder of this paper is organized as follows: Section~\ref{sensing_and_energy_harvesting} comprehensively describes the mechanism of simultaneous sensing and energy harvesting with the resulting concept of energy positive sensing. Task scheduling algorithms for \ac{ehiot} are presented in Section~\ref{Scheduling_in_ehiot} along with energy prediction algorithms, to ensure the perpetual operation of sensor nodes in \ac{ehiot}. Section~\ref{sec:challenges_and_opputunities} critically analyses the challenges and opportunities for devising task scheduling algorithms for energy positive sensing. Future research directions are described in Section~\ref{Future_Research_Directions} and finally, Section~\ref{conclusion} concludes the paper. Table~\ref{table_Nomenclature} provides the nomenclature used in this paper.
\begin{figure}[t!]
\centering
\includegraphics[width=7cm, height=8cm]{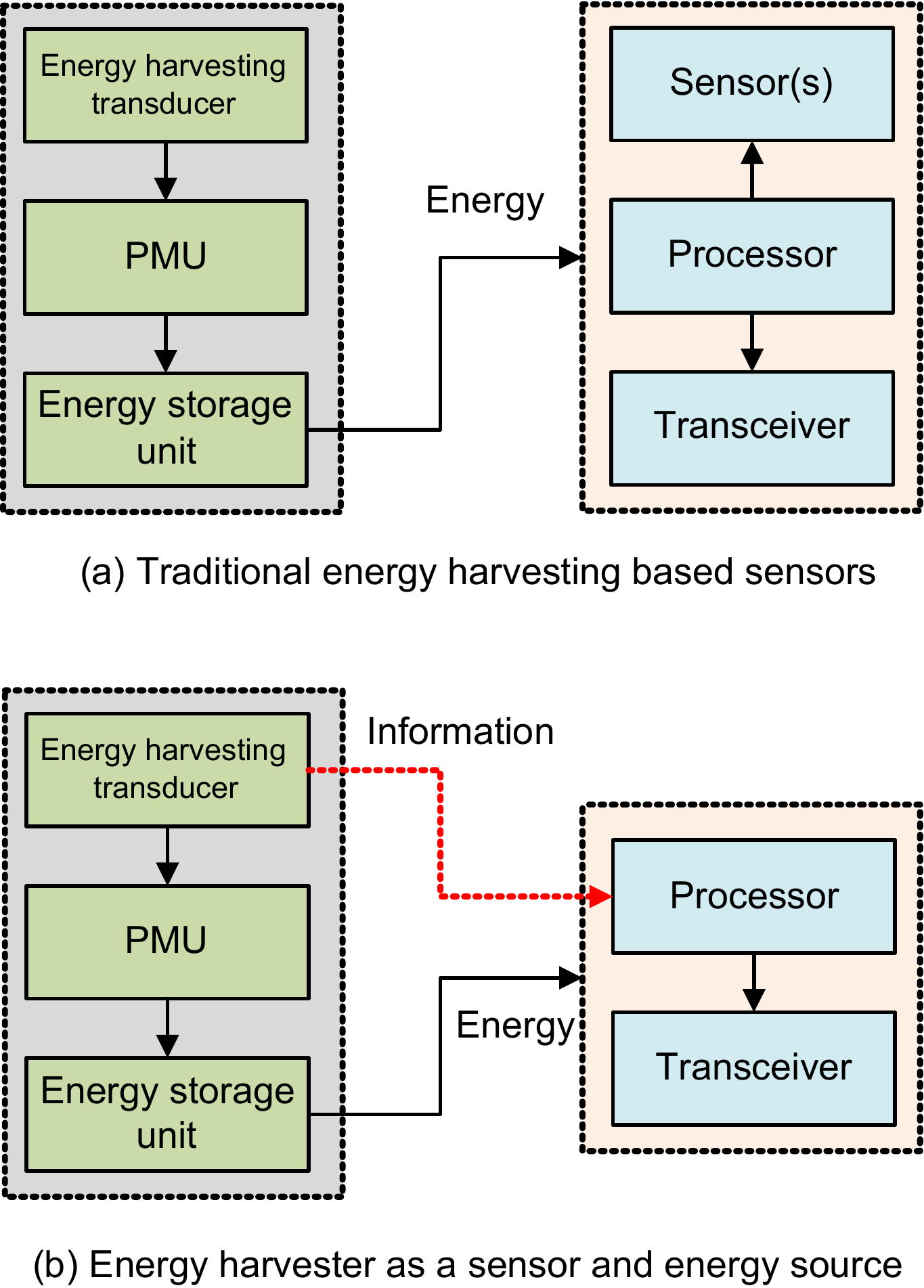}
\caption{Energy harvesters can be used as activity sensors, in addition to the energy scavenging}
\label{fig_sensing}
\end{figure}



\begin{table*}[ht]
\caption{Summary of State-of-the-Art Related to Sensing using Various Energy Harvesters}
\label{table_sensing_papers}
\centering
\begin{tabular}{|c|c|c|c|c|c|c|c|c|}
\hline
\textbf{Year} & \textbf{Reference} & \textbf{Energy harvester} & \textbf{Sensing based on} & \textbf{Target application} & \textbf{Energy stored} & \textbf{Load} & \textbf{Algorithm} & \textbf{Accuracy}\\
 & & & & & & & \textbf{implementation} &\\
\hline\hline
2012 & \cite{martin2012doubledip} & Thermal & $V_{oc}$ \& $V_{cap}$ & Water flow detection & No & No & Offline & --\\
\hline
2013 & \cite{xiang2013powering} & Kinetic & $V_{rec}$ & Airflow speed monitoring & Yes & No & Offline & -- \\
\hline
2014 & \cite{campbell2014energy} & Thermal & $V_{cap}$ & Water and appliance metering & No & No & Offline & --\\
\hline
2015 & \cite{zarepour2015remote} & Thermal & Signal pulse & Chemical reaction detection & No & No & Offline & --\\
\hline
2015 & \cite{khalifa2015step} & Kinetic & $V_{oc}$ & Human step count & No & No & Offline & 96 \%\\
\hline
2015 & \cite{lan2015estimating} & Kinetic & $V_{oc}$ & Calories burnt & No & No & Offline & -- \\
\hline
2015 & \cite{kalantarian2015monitoring} & Kinetic & $V_{oc}$ & Food intake detection & No & No & Offline & 86 \%\\
\hline
2015 & ~\cite{khalifa2015energy} & Kinetic & $V_{oc}$ & Human activity recognition & No & No & Offline & 99 \%\\
\hline
2015 & \cite{khalifa2015pervasive} & Kinetic & $V_{oc}$ & Human activity recognition & No & No & Offline & 83 \% \\
\hline
2016 & \cite{khalifa2016feasibility} & Kinetic & $V_{oc}$ & Hotword detection & No & No & Offline & 85 \%\\
\hline
2016 & \cite{khalifa2016bayesian} & Kinetic & $V_{oc}$ & Human activity recognition & No & No & Offline & 91 \%\\
\hline
2016 & \cite{abadi2016energy} & Kinetic $^a$ & Accelerometer & Train route detection & -- & -- & Offline & 97 \%\\
\hline
2016 & \cite{lan2016transportation} & Kinetic & $V_{oc}$ & Transport mode detection & No & No & Offline & 85 \%\\
\hline
2016 & \cite{blank2016ball} & Kinetic & $V_{oc}$ & Ball impact on racket & No & No & Offline & -- \\
\hline
2016 & \cite{kalantarian2016pedometers} & Kinetic & $V_{cap}$ & Human step count & Yes & No & Offline & 89 \% \\
\hline
2017 & \cite{zarepour2017semon} & Thermal & Signal pulse & Chemical reaction detection & No & No & Offline & --\\
\hline
2017 & \cite{lan2017capsense} & Kinetic & $V_{cap}$ & Human activity recognition & Yes & No & Offline & 96 \%\\
\hline
2017 & \cite{varshney2017battery} & Solar & $V_{oc}$ & Hand gesture recognition & No & Yes & Offline & -- \\
\hline
2017 & \cite{xu2017keh} & Kinetic $^b$ & $V_{oc}$ & Human gait recognition & No & No & Offline & 95 \%\\
\hline
2017 & \cite{pradhan2017rio} & RF & Signal phase & Touch detection & No & No & Offline & --\\
\hline
2017 & \cite{lan2017veh} & Kinetic $^c$ & $V_{R}$ & Voice demodulation & No & Yes & Offline & --\\
\hline
2018 & \cite{lan2018capacitor} & Kinetic & $V_{cap}$ & Human activity recognition & Yes & No & Offline & 95 \%\\
\hline
2018 & \cite{ma2018sehs} & Kinetic & $V_{AC}$ \& $V_{cap}$ & Human gait recognition & Yes & No & Offline & 86 \%\\
\hline
2018 & \cite{khalifa2018harke} & Kinetic & $V_{oc}$ & Human activity recognition & No & No & Offline & 95 \%\\
\hline
2018 & \cite{safaei2018energy} & Kinetic & $V_{oc}$ & Knee surgery monitoring & No & No & Offline & -- \\
\hline
2018 & \cite{wang2018challenge} & RF & Signal phase & Hand gesture sensing & No & No & Offline & --\\
\hline
2019 & \cite{ma2018solargest} & Solar & $V_{oc}$ & Hand gesture recognition & No & No & Offline & 96 \%\\
\hline
2019 & \cite{lan2018entrans} & Kinetic & $V_{cap}$ & Transport mode detection & No & No & Offline & 92 \%\\
\hline
2019 & \cite{xu2019keh} & Kinetic $^b$ & $V_{oc}$ & Human gait recognition & No & No & Offline & 96 \%\\
\hline
2019 & \cite{lin2019h2b} & Kinetic & $V_{oc}$ & Heart beat monitoring & No & No & Offline & --\\
\hline
2019 & \cite{umetsu2019ehaas} & Solar \& Kinetic & $V_{oc}$ & Recognizing places & No & No & Offline & 88 \%\\
\hline
2020 & \cite{sandhu2020ssehkeh} & Kinetic & Current & Human activity recognition & Yes & Yes & Offline & 97 \%\\
\hline
\end{tabular}
\footnotesize{ $^a$ Accelerometer samples are transformed into KEH samples using a mathematical model, $^b$ Employs two KEHs \textit{i.e.,} PEH and EEH}, $^c$ Uses the output power measured across a resistor, which connected between both terminals of the \ac{peh} transducer\\
\end{table*}
 
\section{Simultaneous Sensing and Energy Harvesting}
\label{sensing_and_energy_harvesting}
In addition to energy harvesting, a recent trend is to use the output signals from energy harvesters for extracting the context information, as shown in Fig.~\ref{fig_sensing}(b) compared to using dedicated sensors (e.g. accelerometer, magnetometer, etc.), as depicted in Fig.~\ref{fig_sensing}(a). Thanks to the varying nature of the harvested energy, it contains the context information about the environment in which the energy harvester is deployed. For example, the harvested voltage from \ac{teh} transducer~\cite{martin2012doubledip} contains information about the temperature of the environment.
Similarly, using the harvested energy from \ac{seh} transducer, the indoor and outdoor environments can be differentiated~\cite{umetsu2019ehaas}. Likewise, if the \ac{keh} transducer is placed on the wrist of human body, the output signal provides information about the type of the underlying activity~\cite{khalifa2015energy}. It is based on the phenomenon that \ac{keh} experiences distinct vibration patterns during different human activities (e.g., walking, running, sitting, standing, etc.) of the human body. These different types of activities leave their distinct signatures in the output signal of \ac{keh}. By analysing the output signal from the energy harvester, it is possible to find the type of activity performed. The principal advantage of using the energy harvester for context detection lies in its sensor related power saving, as compared to the conventional activity sensors (such as accelerometers and magnetometers), which operate on the supplied power~\cite{khalifa2018harke} from the ESU, as illustrated in Fig.~\ref{fig_sensing}.

Table~\ref{table_sensing_papers} comprehensively summarizes relevant recent works on using energy harvesters as sensors. This table shows interesting observations form the energy harvesting-based sensing literature.  Firstly, it shows that energy harvesters are used as a source of context information in various applications, including human gait and activity recognition, health and fitness tracking, transport mode detection, as well as in localization and shadow detection (as shown in column 5). Secondly, most of the previous works focus on sensing using the signal received from \ac{seh} or \ac{keh}, as shown in column 3. A limited number of works employ \ac{teh} and \ac{rfeh} as a source of information due to their limited applications, weaker output signal and significant noise component. One interesting point to note is that most of the previous research is focused on extracting context information using \ac{keh}. A small number of previous works consider \ac{seh} and other harvesters as information sources. It can be due to the reason that harvested energy from \ac{seh} is relatively stable as compared to \ac{keh} and does not contain a fluctuating, variable and fine grained pattern compared to the latter. In addition, potentially \ac{seh} can only be used in applications where the context information is contained in changing light conditions and thus it is not applicable in dark environments, such as at night. While most of the previous works use single energy harvester, only one work~\cite{umetsu2019ehaas} considers multi-source energy harvesting based sensing. Authors in~\cite{umetsu2019ehaas} employ \ac{seh} and \ac{keh} to sense different types of environment (i.e., indoor and outdoor), and types of human activities (sitting, walking, running, etc.) respectively.  Multi-source energy harvesting based sensing is useful in applications where the objective is to identify different types of environments that offer changing conditions, such as, light intensity (solar), movement (kinetic) and temperature (thermal). In these applications, one type of energy harvester may not capture the distinct pattern for each type of activity. However, combining the information from multi-source energy harvesters provides rich context information, which can distinguish various types of activities with reasonable accuracy.

\begin{figure}[t!]
\includegraphics[width=7cm, height=4.5cm]{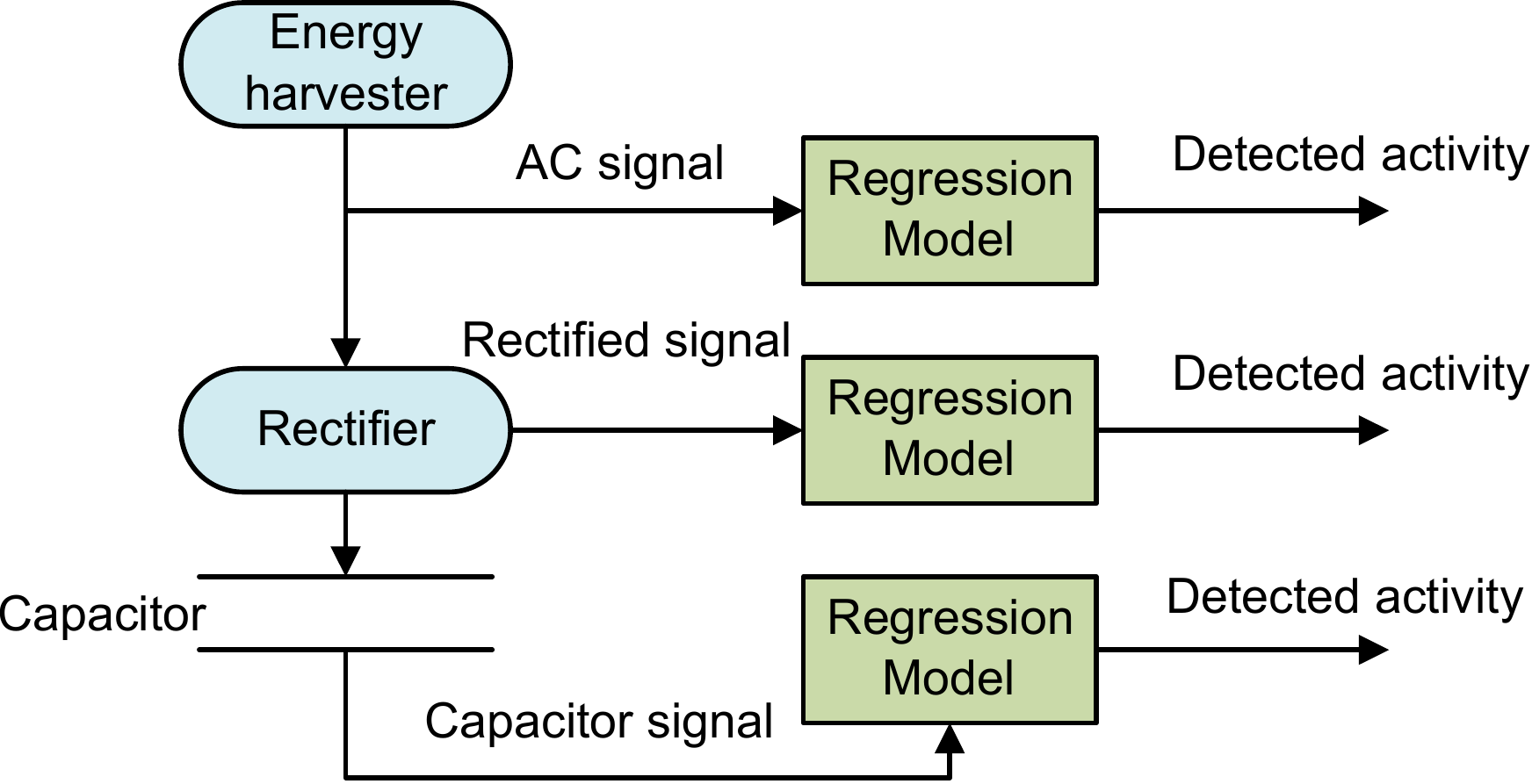}
\caption{Energy harvesters have different types of signals for information extraction, such as AC voltage, rectified voltage and capacitor voltage}
\label{fig_sensing_options}
\vspace{-0.5cm}
\end{figure}

There are different types of signals in the energy harvesting circuit, such as open circuit voltage $V_{oc}$, rectified voltage $V_{rec}$ and capacitor voltage $V_{cap}$, as shown in Fig.~\ref{fig_sensing_options}, that contain the required context information~\cite{sandhu2020ssehkeh}.
From Table~\ref{table_sensing_papers}, we can observe that most previous works use open circuit voltage, while a limited number of works employ the capacitor voltage for context information extraction as presented in column 4. Apart from $V_{oc}$ and $V_{cap}$, there is only one work \cite{xiang2013powering} which uses rectified voltage $V_{rec}$ for sensing.  Additionally, only a limited number of recent works \cite{xiang2013powering, kalantarian2016pedometers, lan2017capsense, lan2018capacitor, ma2018sehs} store the harvested energy in the \ac{esu}, as shown in column 6 of Table~\ref{table_sensing_papers}. This stored harvested energy can be used to power the sensor node or its selected module, depending upon the amount of harvested energy. In addition, the stored energy can be used as an information source~\cite{lan2017capsense, lan2018capacitor, ma2018sehs}. Furthermore, in addition to sensing using energy harvester, none of previous works use the harvested energy to run a system load except~\cite{xiang2013powering}, as illustrated in column 7 of Table~\ref{table_sensing_papers}. Although some previous works use the stored energy for sensing, they still do not employ this energy to power any system load~\cite{lan2017capsense, lan2018capacitor, ma2018sehs}, except~\cite{xiang2013powering}, which uses two separate \ac{keh} transducers one for sensing and the other for energy harvesting. Therefore, it ignores the potential of using both transducers for concurrent sensing as well as energy harvesting. Moreover,~\cite{varshney2017battery} and~\cite{khalifa2016bayesian} use the harvested energy from \ac{seh} and \ac{keh} respectively, to transmit a signal without storing it in an \ac{esu}. The received signal indicates that the (coarse-grained) activity is triggered. It does not extract information from the harvesting signal, instead it uses the harvesting energy to transmit a pulse. The activity is detected on the receiver end by analysing the characteristics of the activity pulse, such as its strength and time duration. On the other hand, when a load is used with the energy harvester, it changes (distorts) the harvesting voltage waveform \cite{ma2018sehs}. As a result, it may affect the information content in the harvested signal. As energy scavenging is generally the major application of energy harvesters, there is only one work~\cite{sandhu2020ssehkeh} that focuses on simultaneous sensing and energy harvesting to detect the activity and to power the dynamic load respectively.

The second last column in Table~\ref{table_sensing_papers} depicts that all of the previous works implement their proposed algorithms offline. This is due to the reason that most of the previous works~\cite{khalifa2018harke,ma2018solargest,umetsu2019ehaas,lan2018entrans} do not harvest energy, instead they use open circuit voltage for context detection. Although some other works~\cite{lan2018capacitor,lan2017capsense,ma2018sehs} store the harvested energy in a capacitor, they manually discharge the capacitor, instead of using a realistic dynamic load.
In addition, the harvested energy from tiny energy harvesters is very small and limited. In addition, due to the non-ideal \ac{esu}s, some of the harvested energy is wasted via leakage. Furthermore, there is also some energy loss during the charging/discharging process in the \ac{esu}. Therefore, it is impractical to implement computationally expensive algorithms on the node, with intermittent and unreliable harvested energy from the environment. However, sophisticated energy harvesting techniques~\cite{sandhu2020optimal} can be incorporated to maximize the harvested energy that can realize the concept of online processing on the sensor node.
As the available energy is precious and limited, it must be used efficiently in applications demanding less energy consumption and operating at a low duty cycle. Furthermore, the harvested energy should match the application's energy consumption profile for \ac{eno} of the system. Finally, most of the previous works attain reasonable context detection accuracy ($>80\%$) for energy harvesting based sensing using well-known machine learning classification algorithms. It proposes that energy harvesting based sensing can be used in place of conventional power hungry activity sensors (such as accelerometers) to save the energy while attaining reasonable context detection accuracy. In order to further enhance the context detection accuracy, deep learning and neural network based models can be employed~\cite{xu2019energy}, which have shown promising results in various applications, such as speech recognition~\cite{zhao2019speech}, face recognition~\cite{li2018occlusion} and natural language processing~\cite{young2018recent}.

In contrast to~\cite{ma2019optimizing}, this survey focuses on the use of energy harvesters as a source of energy and context information simultaneously. This results in full utilisation of energy harvesters, reducing form factor, cost and weight. In addition, we explain the use of harvested energy to acquire the energy harvesting signal (for context detection) leading towards energy positive sensing~\cite{sandhu2020ssehkeh}. In order to provide adequate detail to the reader, we comprehensively explore both of these emerging mechanisms in the following subsections.
\begin{figure*}[t!]
\centering
\includegraphics[width=14cm, height=4cm]{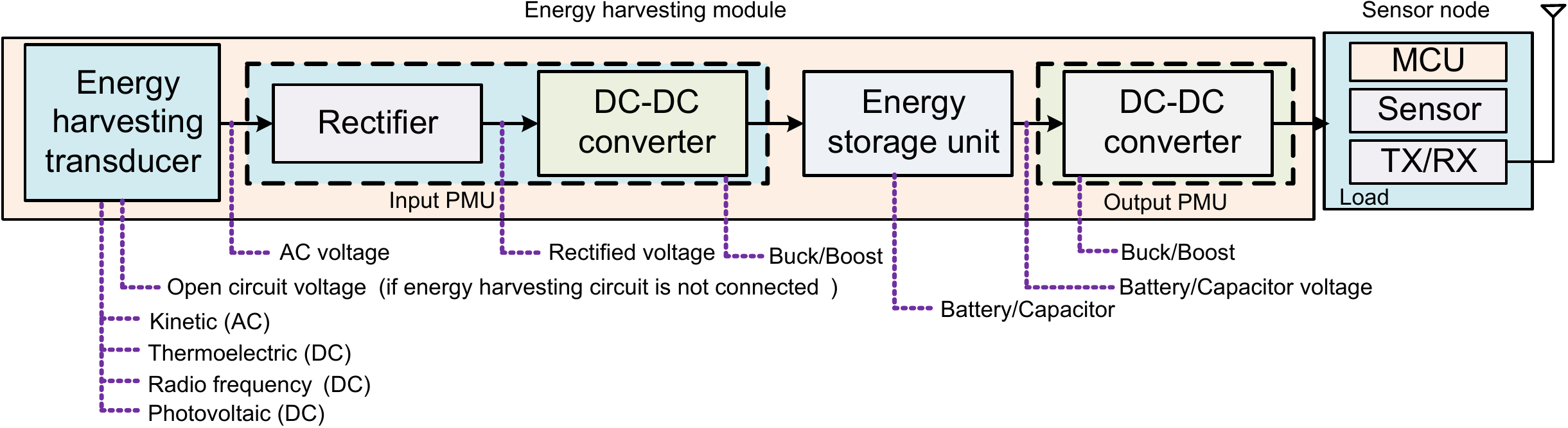}
\caption{General system architecture for energy harvesting to power the sensor nodes in \ac{iot}}
\label{fig_EH_mechanism}
\vspace{-0.5cm}
\end{figure*}

\subsection{Sensing and energy harvesting concurrently}
\begin{figure}[t!]
\includegraphics[width=8.5cm, height=2.5cm]{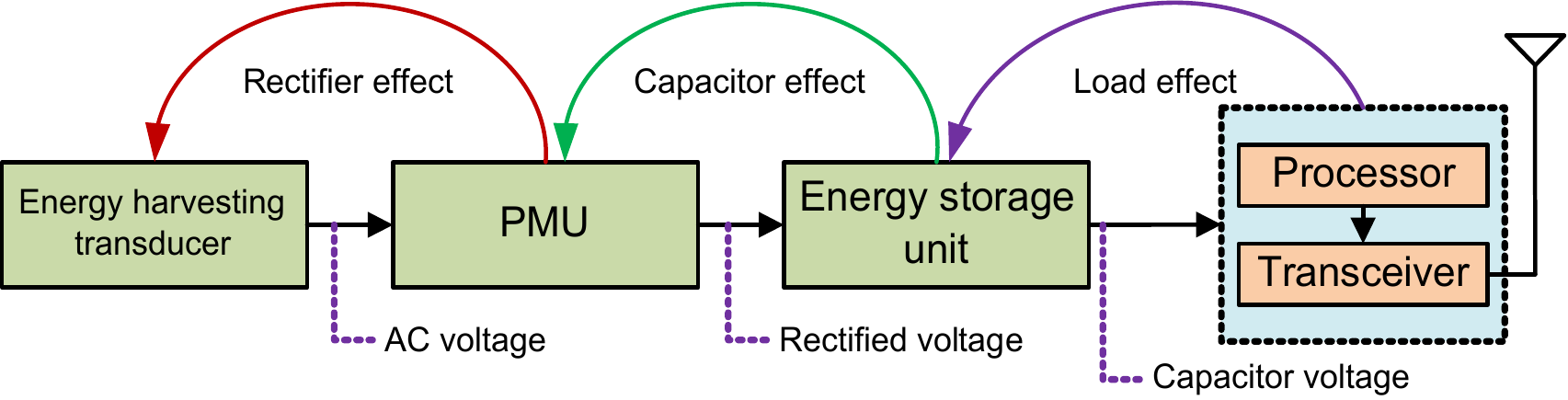}
\caption{Impact of load, stored energy and \ac{pmu} on the harvesting AC voltage of the energy harvesting transducer}
\label{sensing_trend}
\vspace{-0.5cm}
\end{figure}

\begin{figure}[t]
\includegraphics[width=8.5cm, height=4cm]{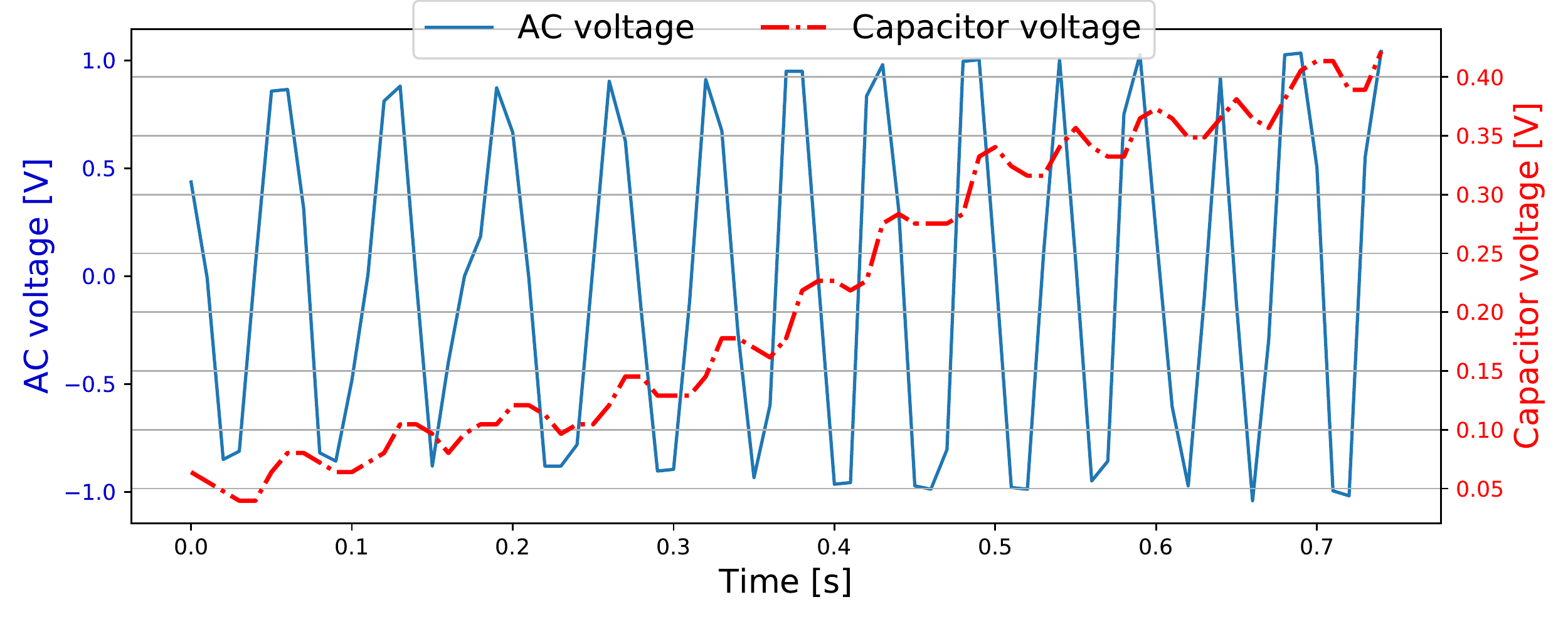}\vspace{-0.25cm}
\caption{The harvesting AC voltage increases with the rise in \ac{keh} capacitor voltage}
\label{fig_sehs_signal_distortion}
\vspace{-0.5cm}
\end{figure}

Energy harvesters can be used as sensor as well as source of energy to detect the underlying activity and to power the sensor nodes. The generated energy from the energy harvesting transducer is stored in an \ac{esu}, which is later used to run a system load. Fig.~\ref{fig_EH_mechanism} highlights the general energy harvesting mechanism using a DC-DC boost converter~\cite{geissdoerfer2019shepherd} in the \ac{pmu}\footnote{In the literature, PMU is also called as Energy Management Unit (EMU)}. The harvested energy is first stored in a capacitor and then, it is used to run hardware device. It is evident from Table~\ref{table_sensing_papers} that most of the previous work is focused on using AC or capacitor voltages for extracting the activity information from the energy harvester. Although it is a widely used practice in the literature, it is practically inefficient for real-world scenarios. This is because energy harvesters are not being used as a source of energy in the open circuit configuration. Similarly, the previous schemes that employ the capacitor voltage for sensing~\cite{lan2018capacitor} manually discharge the capacitor, instead of using a real practical system load (such as a sensor node). Fig.~\ref{sensing_trend} depicts the basic building blocks of an energy harvesting circuit, which include a transducer, \ac{pmu} and an \ac{esu}. When a load is connected to the energy harvesting circuit, it modifies the shape of the input AC as well as capacitor voltages. It is due to the reason that AC voltage of the transducer depends on the capacitor voltage as:
\begin{equation}
    V_{AC} = V_{cap}+V_{pmu}
    \label{eq:voltage_ac_cap_pmu}
\end{equation}
where, $V_{pmu}$ is typically equal to the voltage drop across two diodes in the bridge rectifier. The shape of the harvesting AC voltage depends upon the capacitor voltage, which may impact its embedded information content, compared to the original open circuit AC voltage. 
Similarly, when a load is connected to the capacitor, it discharges the capacitor after irregular time intervals, instead of fixed manual discharging~\cite{lan2018capacitor}. The reason behind it lies in the distinct vibration pattern, for example in \ac{keh}, during various activities, which produces different voltage levels at the output of the transducer; thus charging the capacitor at a different rate. 
Therefore, the harvested energy (stored in the capacitor) distorts the generated AC signal according to Eq.~\ref{eq:voltage_ac_cap_pmu}, if energy harvesting and sensing is performed using the same \ac{keh}. Therefore, the authors in~\cite{ma2018sehs} employ two \ac{keh} transducers to procure accurate context information in the presence of a distorted AC signal. They design a hardware prototype which is embedded in a shoe and contains two \ac{keh} transducers mounted in the front and rear of a shoe.
The harvested AC voltage increases in amplitude due to the rise in the capacitor voltage, as depicted in Fig.~\ref{fig_sehs_signal_distortion}. It is due to the fact that the charging current decreases when the capacitor is being charged to a higher voltage level. Furthermore, \ac{keh} has a significantly high internal resistance of the order of M$\Omega$~\cite{huang2016battery, xiang2013powering} and the output voltage is determined by the load resistance as well as the internal resistance of \ac{peh}~\cite{huang2016battery, sodano2004estimation}. As the current flow is decreased, the voltage on the internal resistance also decreases, enhancing the corresponding output voltage. This modified (and distorted) AC signal may have less information content as compared to the original AC signal. The authors in~\cite{ma2018sehs} propose a filtering algorithm to remove the effect of capacitor voltage on the harvesting AC voltage to enhance the gait recognition accuracy. However, this filtering algorithm is difficult to implement on miniaturized and resource constrained sensor nodes, in real practical environments with a limited and time-varying supply of harvested energy.

Therefore, more sophisticated algorithms are needed to extract useful information from the harvesting signals in the presence of an energy harvesting circuit, while powering a realistic intermittently-powered load. This intermittent operation is due to the use of capacitors as \ac{esu}, which store a small amount of energy to run at most one atomic task, in contrast to  batteries which can power nodes continuously for a longer duration. One drawback of employing capacitor voltage for activity detection is that the harvested energy, in some applications, is not enough to quickly charge the capacitor. In other words, sometimes, it takes longer to charge the capacitor up to a certain required load voltage (especially under lower vibrations), which may introduce delay in activity detection. One solution is to employ a smaller capacitor, that can be charged quickly at the cost of smaller energy burst for the load which is enough to run the node for executing at least one atomic task. In summary, it is important to devise a sensing mechanism, which provides activity information using the altered harvested signal in the presence of a capacitor and system load. This configuration will eventually let the energy harvester to work as both source of energy and information simultaneously, leading towards autonomous sensors in \ac{ehiot}.

Authors in~\cite{ma2018sehs} do not employ the harvested energy to power a system load, instead, they discharge the capacitors manually, which are recharged again with the harvested energy from the transducers. Sandhu et al.~\cite{sandhu2020ssehkeh} employ energy harvesters as sensors and source of energy simultaneously. They use two \ac{leds} as load, which are powered using the stored harvested energy from the \ac{keh} transducer. In addition, they explore various sensing points in the energy harvesting circuit, that offer two types of signals: current and voltage. They evaluate the sensing potential of various available signals in the energy harvesting circuit and show that the harvesting current signal offers highest activity classification accuracy, while powering a dynamic load.

\subsection{Energy positive sensing}
\label{energy_positive_sensing}
Simultaneous sensing and energy harvesting enables \textit{energy positive sensing} -- an important and emerging class of sensors, which harvest higher energy than required for the signal acquisition. This additional harvested energy can be used to power other hardware modules on the node, moving towards \ac{eno}.
Eventually, it leads towards the design and implementation of autonomous sensor nodes in \ac{ehiot}, that solely operate using the harvested energy and employ the energy harvesting transducers as sensors for a theoretically indefinite lifetime. Sandhu et al.~\cite{sandhu2020ssehkeh} are the first to explore the energy positive sensing concept in transport mode application.

The energy harvesting signal needs to be digitized before processing by the embedded device. In digitizing the harvesting signal, an \ac{adc} consumes, what we call, signal acquisition power~\cite{sandhu2020ssehkeh}. This signal acquisition power can be lower or higher than the harvested power, depending upon the characteristics of the transducer and environment in which it is employed. The Acquisition Power Ratio (APR)~\cite{sandhu2020ssehkeh} is quantitatively defined as:
\begin{equation}
    APR = \frac{P_{har}}{P_{acq}}
    \label{eq:apr_energy_positive sensing}
\end{equation}
Depending on the amount of harvested energy, sensor nodes fall into three major categories.
\begin{itemize}
    \item Energy positive sensing (if $APR>1$)
    \item Energy neutral sensing (if $APR=1$)
    \item Energy negative sensing (if $APR<1$)
\end{itemize}
Energy negative sensors have lower harvested energy than required by the \ac{adc} to digitize the sensing signal and thus need an external energy source to power the nodes. Energy neutral sensing lies at the boundary between energy negative sensing and energy positive sensing, and has less practical significance. Energy positive sensor is an important and emerging class of sensors, which harvest higher energy than required for the signal acquisition, which can be used to power other hardware modules on the node.

Sandhu et al.~\cite{sandhu2020ssehkeh} extensively explore energy positive sensing mechanism using \ac{keh} hardware prototypes. They use transport mode detection as a case study and employ two different designs for the energy harvesting circuit: with and without DC-DC boost converter. They collect data from six types of transport modes, including ferry, train, bus, car, tricycle and pedestrian movement. In addition to the harvesting signals, they use the harvested energy to power a dynamic load. Their results show that, on average, DC-DC converter based energy harvesting circuits harvest more energy than required for signal acquisition. Thus, it can lead towards the aforementioned energy positive sensing paradigm, resulting in self-powered and autonomous operation of batteryless sensor nodes in \ac{ehiot}.

\subsection{Summary and insights}
Energy harvesters serve the dual purpose of energy generation as well as context detection, addressing the requirement of conventional activity sensors, such as accelerometers, magnetometers and gyroscopes. This mechanism saves energy that would otherwise be consumed in powering the conventional activity sensors. Most of the previous works employ \ac{keh} transducers for context detection due to their wide applicability and strong output signal. In order to further save the energy consumption, the stored energy in the \ac{esu} can be sampled at a lower sampling rate for context detection, at the cost of higher latency. However, the harvested energy is still not sufficient to continuously power the sensor node using miniaturized energy harvesters, such as placed on the wrist in human fitness tracking applications. This opens the door for energy management algorithms, that schedule the execution of tasks on the node, according to the harvested energy profile, to prolong the operational lifetime of the system. Considering the energy harvesting based sensing mechanism, in the next section we comprehensively survey various task scheduling algorithms that manage the time-varying and limited harvested energy on the miniaturized sensor nodes in \ac{ehiot} to enhance their operational lifetime.

\section{Task Scheduling in \ac{ehiot}}
\label{Scheduling_in_ehiot}
The harvested energy from miniaturized sensors in \ac{ehiot} is not sufficient to power the sensing hardware continuously without any interruption. Therefore, task scheduling based energy management schemes are required in \ac{ehiot} that ensure the efficient utilization of harvested energy. This section surveys existing task scheduling schemes for \ac{ehiot}, with a particular focus on schemes that have the potential to support sensing modalities that harvest energy and sense simultaneously. 
These algorithms ensure optimal utilization of harvested energy to extend system lifetime as well to provide highest activity detection/monitoring performance.

\begin{figure}[t!]
\centering
\includegraphics[width=8cm, height=6cm]{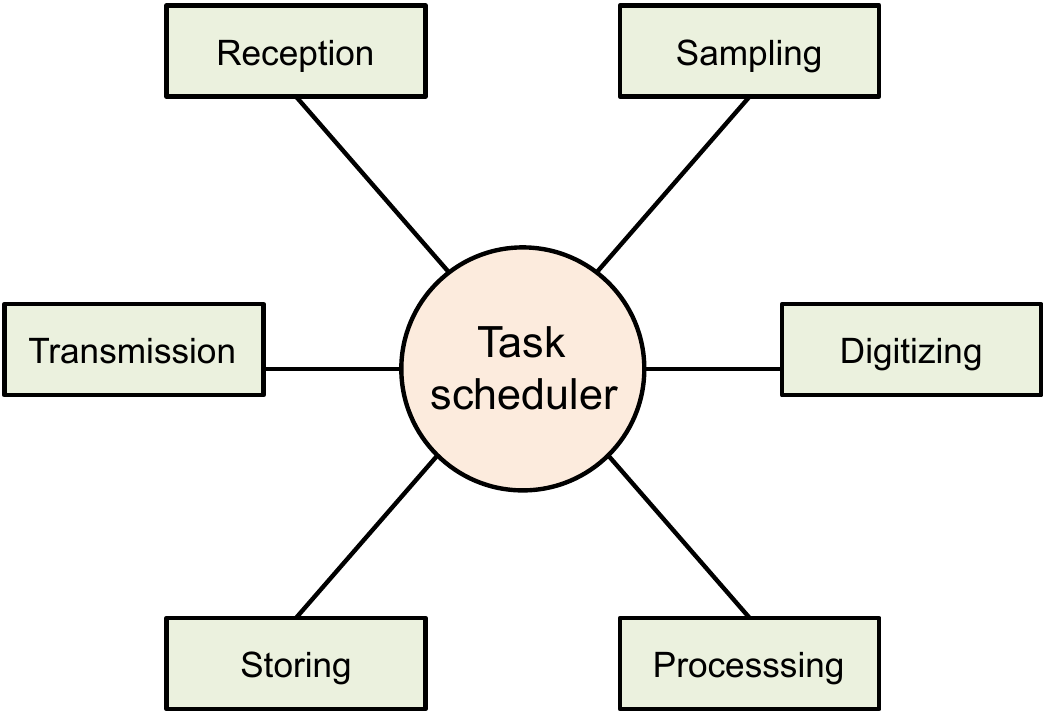}
\caption{The task scheduler governs the execution of various tasks on the sensor node in \ac{ehiot}}
\label{fig:types_of_tasks}
\vspace{-0.5cm}
\end{figure}

\begin{figure*}[ht]
\centering
\includegraphics[width=15cm, height=6cm]{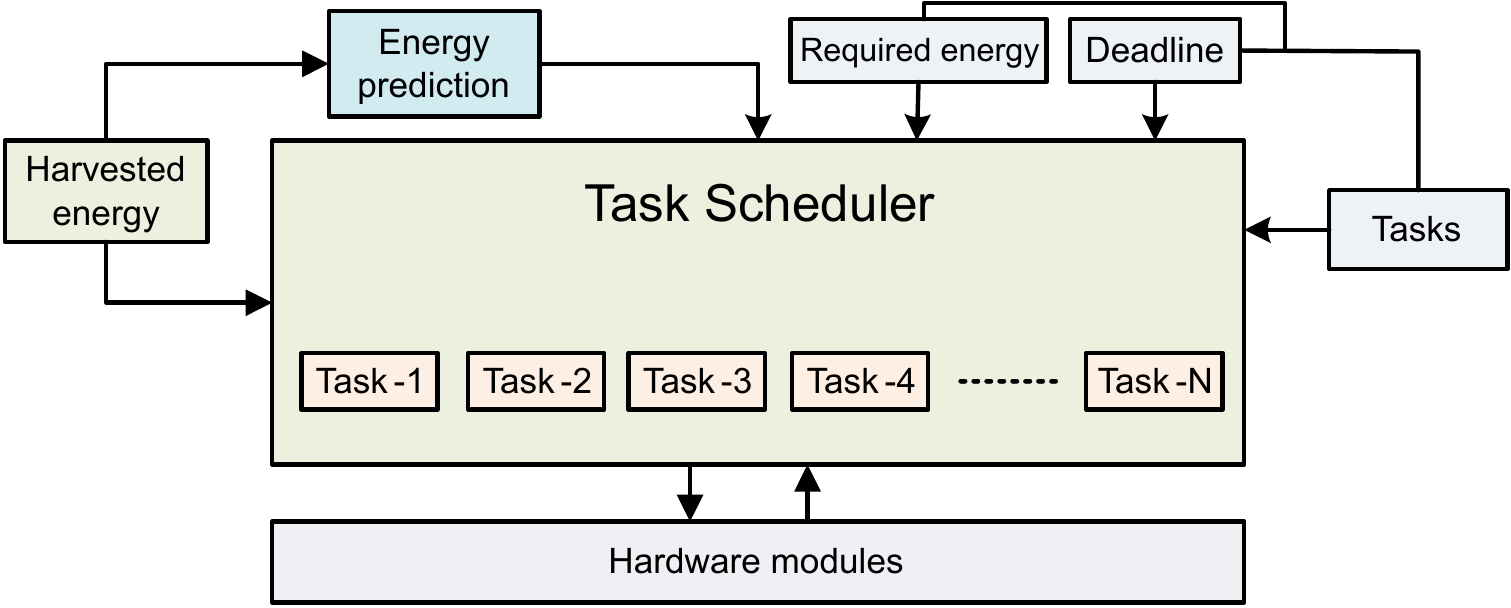}
\caption{The task scheduler schedules the tasks based on different parameters including task's energy consumption, deadline and harvested energy}
\label{fig_scheduling}
\vspace{-0.5cm}
\end{figure*}

Task scheduling schemes are devised to manage the execution of tasks on the node to maximize the performance in terms of activity monitoring within the limited available energy budget. This means that the objective of task scheduling schemes is to minimize the energy consumption and deadline miss ratio, maximize the harvested energy and ensure the operation of sensor node for extended time period with higher accuracy of activity monitoring/detection according to the type of application. The most common types of tasks on the node include sampling of information/signal, digitizing, processing, storing, data transmission and reception, as displayed in Fig.~\ref{fig:types_of_tasks}. According to the type of application, the tasks are scheduled by the task scheduler for the execution by the processor on the sensor node at different frequency. The task scheduler takes various parameters into account, while deciding the execution of upcoming tasks on the node, as depicted in Fig.~\ref{fig_scheduling}. The figure shows that the task scheduler considers the energy budget, task deadline, predicted energy and type of task(s), while scheduling the execution of tasks on the node. If the available energy is insufficient, energy-intensive tasks can be decomposed into smaller subtasks, which consume less energy during their execution. In the previous literature, the task scheduling techniques are based on the following key principles:
\begin{itemize}
    \item Tasks are queued according to their priority and order of their deadlines.
    \item As long as sufficient energy is available, a high priority task is executed depending on the amount of harvested energy.
    \item If the harvested energy is not enough to run a high priority task, the next task in the queue is executed.
    \item Low priority tasks are executed, if there is energy left after the execution of high priority tasks.
    \item If sufficient energy is not available, the tasks are delayed until their deadlines, to allow the transducer to harvest energy.
    \item If a high priority task arrives during the execution of a low priority task, the low priority task is pre-empted (according to the type of application), to execute the high priority task.
    \item Tasks are scheduled according to the available energy, predicted harvested energy, required energy and deadlines of tasks.
    \item Tasks can be broken down into smaller subtasks, which consume small amount of energy and take less time during their execution. Later, these subtasks can be combined, according to their similarity, to reduce the energy consumption in frequent switching of the hardware from idle mode to active state.
\end{itemize}

\begin{figure*}[t]
\centering
\includegraphics[width=13cm, height=8cm]{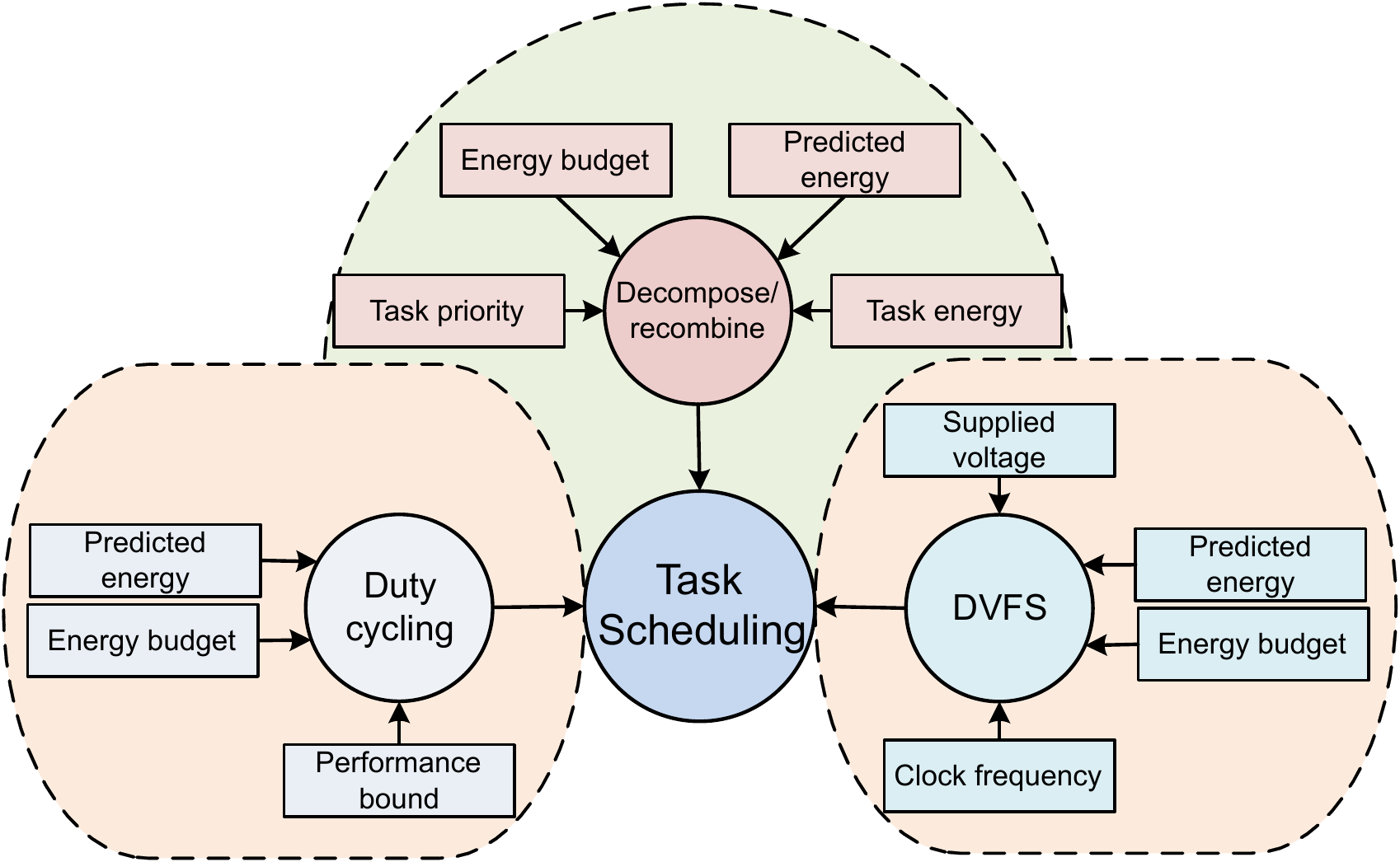}
\caption{Different mechanisms for scheduling the tasks on the sensor node}
\label{fig_task_scheduling}
\vspace{-0.5cm}
\end{figure*}

Fig.~\ref{fig_task_scheduling} displays a summary of major task scheduling algorithms in \ac{ehiot}. In a nutshell, there are three major strategies employed for devising a task scheduling algorithm for sensor nodes in \ac{ehiot}:
\begin{itemize}
	\item \ac{dvfs}
	\item Decomposing and combining of tasks
	\item Duty cycling
\end{itemize}
Table~\ref{table_surveyed_papers} provides a brief overview of different task scheduling schemes in the literature which use one or more of these strategies.
The goal of the scheduling schemes is to run the sensor node within the limited available energy budget to meet tasks deadlines, enhance the performance in terms of activity detection and to achieve \ac{eno}.
Table~\ref{table_surveyed_papers} comprehensively describes the state-of-the-art related to task scheduling schemes, their performance metrics and their evaluation methods (i.e., simulation or hardware implementation). Some of the previous works employ \ac{dvfs} to manage the harvested energy and schedule the tasks on the node by dynamically adjusting the voltage and frequency in \ac{ehiot}, as discussed in detail in Section~\ref{DVSF}. The nodes are equipped with energy harvesters that generate energy to run the tasks, and the node switching is controlled using \ac{dvfs}. Furthermore, the voltage level can be adjusted to power the active hardware module only, instead of the whole node, to minimize the energy consumption. Secondly, larger tasks can be decomposed into smaller subtasks that consume less energy during their execution, as described in Section~\ref{Decomposing_and_combining_tasks}. Identical tasks can be grouped together to save energy that would otherwise be consumed in repeated switching of the node's hardware. For example, instead of transmitting two data packets separately, they can be merged to save the energy required to initialize the radio transceiver. Duty cycling is another task scheduling mechanism that allows controlling the consumed energy by the nodes when they are not performing any useful operation, as discussed in Section~\ref{subsection:duty_cycling}. The nodes in \ac{ehiot} are turned on to execute the tasks according to the harvested energy, type of the task and energy demand. This results in lower energy consumption as nodes are switched to low energy modes during their idle time slots. 

The predicted harvested energy plays an important role in scheduling the tasks on the node, when the available energy is scarce. Table~\ref{table_surveyed_papers} shows that there are two main approaches for considering the predicted harvested energy in the previous works: devising a model for energy computation (C) or using previous available algorithms (R). Most of the previous works devise an energy prediction model, which computes the future harvested energy using the previous harvested energy samples and environmental parameters, as discussed in Section~\ref{subsec:predicting_harvested_energy_SEH_IoT} in detail. Low priority tasks that require higher energy can be delayed if there is a prediction of higher harvested energy in the future. This allows to utilize the available limited harvested energy for running the high priority (and low energy) tasks without violating their deadlines. 
Another objective of scheduling schemes is to meet tasks deadline.
High priority tasks are executed ahead of low priority tasks to meet the \ac{qos} requirements. Depending on the application, the low priority tasks can be pre-empted during their execution, when high priority tasks arrive at the task scheduler. 
Finally, tasks are executed within the available energy budget to achieve \ac{eno} on the sensor node.
Tasks are scheduled according to the available energy and required energy to run the tasks, as shown in Table~\ref{table_surveyed_papers}. If the available energy is lower compared to the requirement of a high energy task, a low energy task is executed, even though it has lower priority, to utilize the valuable harvested energy for useful operation. As illustrated in Table~\ref{table_surveyed_papers}, the task scheduling algorithms can be evaluated using two approaches: performing simulations or implementing in a real-world scenario. 

Table~\ref{table_surveyed_papers} portrays that most of the previous works validate their algorithms using simulations instead of real hardware implementation. The last column of Table~\ref{table_surveyed_papers} depicts the difficulty level in implementing the given task scheduling algorithms on the energy harvesting based sensing device. \ac{dvfs} based algorithms are highly difficult to implement on the tiny and miniaturized sensor nodes, due to the stringent requirement of complex circuitry that provides various voltage levels for different components of the node. Similarly, the algorithms that are validated in simulations need significant effort to be implemented on the real energy harvesting based sensor, due to different hardware design for this new class of sensors that run intermittently without conventional batteries. Furthermore, the algorithms that decompose and recombine tasks may not be suitable due to limited amount of harvested energy that can be used to run, at most, one atomic task at a time.
However, duty cycling based scheduling mechanisms that are implemented on real hardware platforms can be implemented on the energy harvesting sensor. It is due to their ease of implementation and compatibility with intermittently powered sensor nodes. However, it needs significant revision as the energy harvester acts as a sensor and source of energy simultaneously without conventional \ac{esu}; therefore, we place them in medium difficulty level, as displayed in Table~\ref{table_surveyed_papers}. 

Most interestingly, none of the previous task scheduling algorithms employs energy harvesters as activity/motion sensors. All existing task scheduling schemes employ energy harvesters as a source of power and use conventional sensor modules for the target application, consuming considerable energy during their operation. This puts further pressure on the limited stored harvested energy, due to the requirement to power both the sensor as well as the processing unit, which reduces the overall system lifetime. Therefore, there is a potential to devise dedicated task scheduling algorithms that employ energy harvesters as sensors and a source of energy simultaneously~\cite{sandhu2020ssehkeh}. The objective of these task scheduling schemes is to maximize the lifetime of sensor nodes as well as to achieve the required performance level, using the intermittent and limited harvested energy. Interestingly, this harvested energy has some synchronization with the underlying activity and thus energy is harvested when an activity is performed. Later, this harvested energy is employed to sample the energy harvesting signal to identify the type of activity. This leads towards the possibility of autonomous and self-powered \ac{ehiot} systems, due to the elimination of sensor-related energy consumption by exploiting the embedded information in the energy harvesting signals for context detection applications.

\subsection{Dynamic voltage and frequency scaling}
\label{DVSF}
The power consumption of the sensor hardware depends on the supplied voltage (as well as the current flow) and the operating frequency (i.e., the clock frequency). Therefore, both of these parameters can be adjusted in real-time to optimize the power consumption of the sensor node. The objective of \ac{dvfs} is to maximize the performance given the limited energy budget or to minimize the energy consumption under a performance bound. Authors in~\cite{allavena2001scheduling} present a task scheduling technique using \ac{dvfs} for rechargeable real-time systems. The tasks are grouped according to their deadlines and are executed if the available energy is higher than a certain minimum threshold. Liu et al.~\cite{liu2008energy} propose a scheduling mechanism which slows the processing of tasks according to the available as well as the predicted harvested energy in the future in \ac{seh} based systems. The processor frequency depends on the deadline of tasks and the available energy budget. The task is executed at increasing speed as it approaches its deadline. Authors in~\cite{liu2009adaptive} decouple the energy and timing constraints to simplify the scheduling problem. In addition to the use of \ac{dvfs}, they delay the execution of high energy tasks, if the stored energy is not enough, until sufficient energy is harvested for the execution of tasks on the node.

A \ac{dvfs} algorithm using load matching and considering battery charge/discharge overhead is presented in~\cite{liu2010load}. The load matching is important for maximum power output from the energy harvesting transducer. Since, some amount of the harvested energy is wasted in battery leakage, the algorithm in~\cite{liu2010load} decides between the stored energy or direct use of the harvested energy in running the sensor node. This saves the energy that would otherwise be wasted due to the non-ideal \ac{esu}. A \ac{dvfs} based task scheduling algorithm for structural health monitoring is presented in~\cite{ravinagarajan2010dvfs}. In this scheme, both periodic as well as sporadic tasks are scheduled using a linear regression based algorithm. The sporadic tasks are executed according to their energy and timing constraints to maximize the \ac{qos}. Liu et al.~\cite{liu2012harvesting} combine the static and adaptive scheduling techniques with \ac{dvfs} to attain high performance with energy and timing constraints. Their algorithm adaptively schedules the tasks when there is a prediction of energy overflow, to achieve the maximum benefit from the harvested energy. It also considers the future harvested energy during the scheduling of tasks.
Lin et al.~\cite{lin2013framework} propose a task scheduling framework based on \ac{dvfs}. They track the \ac{mpp} of the \ac{seh} to harvest maximum energy from a solar cell. The tasks are scheduled according to the predicted energy, available energy budget and the deadline of tasks to minimize the task drop ratio.
Tan et al.~\cite{tan2016dynamic} model the energy harvesting system as an energy model, a task model and a resource model and present a task scheduling algorithm based on \ac{dvfs}. The method combines the free dispersed time slots together to harvest maximum energy for running the tasks, which results in the execution of a higher number of tasks within the limited energy budget.

\ac{dvfs} algorithms are not suitable for scheduling tasks on energy harvesting based sensors due to the resource-constrained hardware. As the energy harvesting circuit is intentionally kept simple (to avoid energy losses), it can not provide various voltage levels to execute tasks on the sensor node. In addition, the miniaturized sensor nodes can easily operate using a fixed voltage level, while drawing different amount of current during various operations, which reduces the power (and energy) wastage. Therefore, alternate task scheduling algorithms are required that work seamlessly without additional overhead in terms of energy and resources, to minimize the energy consumption on energy-constrained sensor nodes.

\subsection{Decomposing and combining of tasks}
\label{Decomposing_and_combining_tasks}
\begin{figure}[t!]
\centering
\includegraphics[width=8cm, height=5cm]{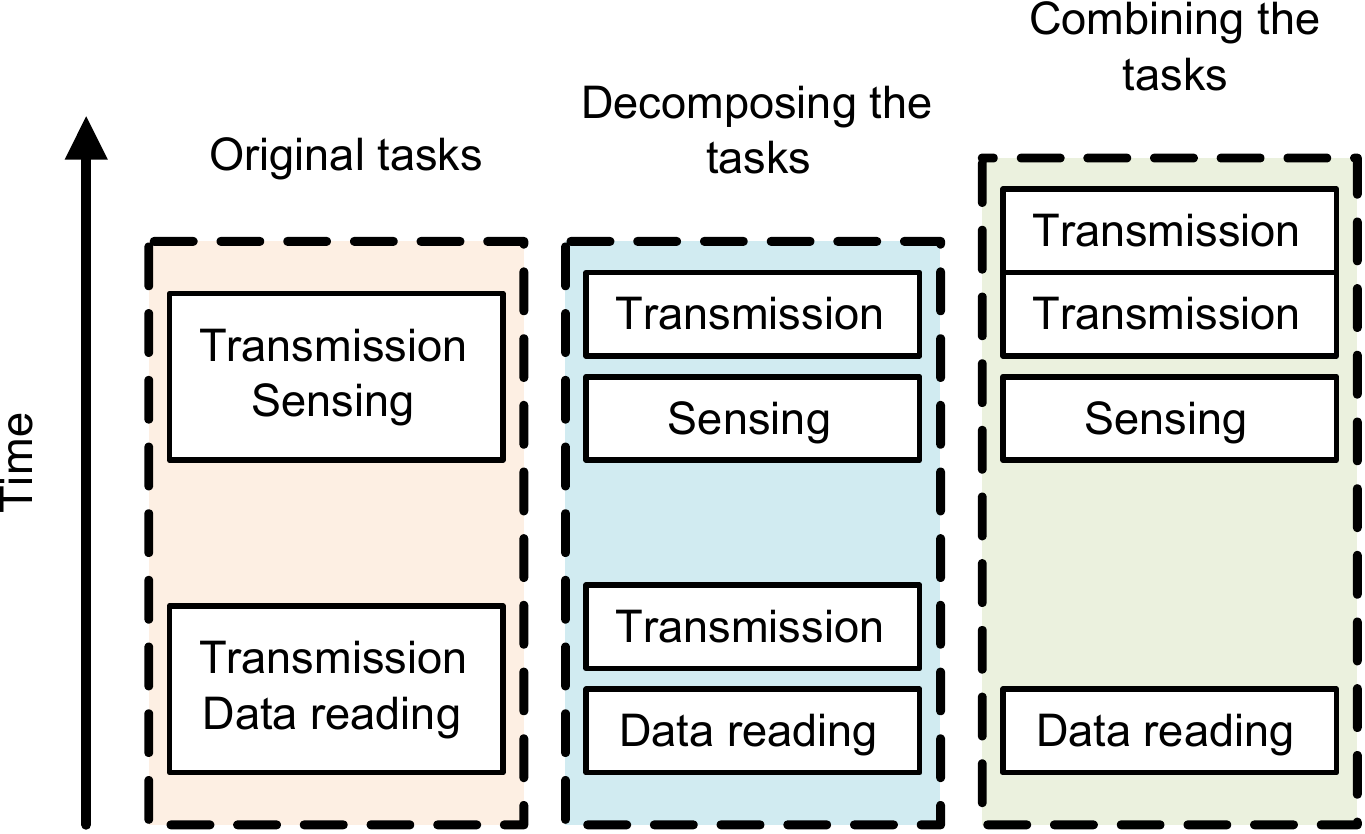}
\caption{Illustration of decomposition and combining of tasks to reduce the energy consumption}
\label{fig_decomp}
\end{figure}

As the harvested energy in \ac{ehiot} is limited, it is difficult to run the power intensive tasks continuously. Nevertheless, if such tasks are executed, the remaining energy may not be sufficient to run the future incoming time-critical tasks. A promising solution of this issue is to decompose the energy-intensive tasks into multiple subtasks which demand lower energy during their operation. For example, the task of transmitting the sensed data can be decomposed into two separate subtasks of sensing and data transmission. Similarly, the transmission of a saved parameter (in memory) can be decomposed into the separate subtasks of reading the attribute from memory and data transmission. In order to minimize energy consumption, the two transmission subtasks can be combined together by grouping the data from these subtasks and transmitting it in one data packet, as shown in Fig.~\ref{fig_decomp}~\cite{zhu2012deos}. Zhu et al.~\cite{zhu2012deos}, decompose larger tasks into multiple subtasks depending upon their ability to recombine with other smaller tasks to save energy.

In summary, the decomposing and combining technique in~\cite{zhu2012deos} consists of the following four phases:
\subsubsection{Decomposition}
This phase decomposes the energy-intensive tasks into multiple subtasks depending upon their ability to group with other subtasks to conserve energy. As the harvested energy is not sufficient to run the high power tasks continuously, subtasks can be executed with the limited available energy budget.
\subsubsection{Combining}
This phase combines multiple subtasks that can be executed on the same processor to minimize the energy consumption. In addition, some tasks can be executed concurrently to reduce the idle time of the processor. For example, the tasks of sensing and reading a value from memory can be executed simultaneously as they utilize different resources of the node, depending on the harvested energy. The advantage of the concurrent execution is the reduced delay and smaller latency in data transmission. However, it also demands higher energy which is available only once in a while in \ac{ehiot}.
\subsubsection{Admission control}
In this phase, the tasks are filtered according to their priority and energy consumption during their execution. Although tasks are combined to save energy, generally, the harvested energy is not enough to run all ready-to-execute tasks. Therefore, an admission controller further filters the tasks based on the following task-specific parameters:
\begin{itemize}
      \item Priority of tasks
      \item Available harvested energy
      \item Energy consumption of tasks
 \end{itemize}
Task priority is important in all applications and in particular for time-critical real-world scenarios. Depending upon the application, the task's deadline is further categorized into two types, i.e. soft deadline and hard deadline. In general, soft deadlines are less critical as compared to hard deadlines and their violation does not harm the functioning of the system. On the contrary, hard deadlines are essential to be respected in all circumstances, which create major loss if ignored or violated~\cite{buttazzo2011hard}. Therefore, the admission controller arranges the tasks according to their priorities and deadlines. In addition to the task priority, the available energy and task energy consumption are also important factors which help in deciding the execution of tasks. If the available energy is lower than the required for the task execution, the task can be delayed so that sufficient energy is accumulated in the \ac{esu} (from energy harvesters) to serve the task successfully.

\subsubsection{Optimization}
This phase optimizes the execution of tasks on the basis of the following attributes:
\begin{itemize}
  \item The additional available energy (available after executing the current task)
  \item The required number of executions of each task
  \item Energy consumption of each task
\end{itemize}
The optimization phase further filters the tasks in order to use the harvested energy efficiently. The additional available energy is important to decide about the execution of future time-critical tasks. In order to avoid deadline violations in the future, a certain minimum amount of energy must be available during all time slots, to serve the future tasks that possess hard deadlines.
Authors in~\cite{zhu2012deos} evaluate their scheme using \ac{seh} as a source of energy. Results show that their technique executes more tasks with fewer missed deadlines as compared to the previous schemes which do not employ the decomposing/combining strategy for energy-intensive tasks.

In addition to decomposing and combining of tasks, a \ac{mdp} model is proposed in~\cite{rao2015optimal} to schedule the tasks on the node. It proposes a dynamic optimization model based on \ac{mdp} to schedule the tasks taking into account their deadlines, energy consumption, and available harvested energy. The decomposed subtasks can be combined together for concurrent execution. It also proposes a less complex greedy scheduling policy which can be implemented on the resource constrained sensor nodes and it consumes less energy as compared to the original model. The simulation results indicate that the proposed algorithm executes higher number of tasks within the same energy budget compared to the previous schemes.

As the harvested energy in energy harvesting-based sensors is limited, they can essentially perform only one atomic task at a time. In the particular case of batteryless sensors, the available energy burst (in the \ac{esu}) can not be employed to execute multiple tasks simultaneously. Therefore, combining the tasks to reduce energy consumption is not appropriate solution for energy harvesting based sensors. As a result, there is a potential of alternate task scheduling algorithms that can provide \ac{eno} on the sensor nodes with limited energy budget.

\subsection{Duty cycling}
\label{subsection:duty_cycling}
One of the most familiar and common methods for minimizing the energy consumption in \ac{ehiot} is to use a sleep/awake mechanism. When a node is not executing any useful operation, it is switched to the sleep mode, which reduces its energy consumption~\cite{7422797}. In traditional \ac{iot}, most of the protocols devise a duty cycling mechanism based on the number of tasks, their energy consumption and remaining energy of the node. However, these methods are not suitable for \ac{ehiot}, due to the variable nature of incoming harvested energy. In \ac{ehiot}, the harvested energy is varying with the environmental conditions due to the intermittent nature of ambient energy harvesting sources~\cite{Adu-Manu:2018:EWS:3203093.3183338}. Therefore, the duty cycle of the node depends on the incoming harvested energy to efficiently utilize the energy in executing tasks on the node. In addition to the current harvested energy, the sleep duration of the node also depends on the energy to be harvested in future to proactively plan the consumption of incoming energy for required operations. Therefore, we classify the duty cycling mechanisms into the aforementioned two categories and extensively explore their operations in the following subsections.

\subsubsection{Duty cycling depending on the harvested energy}
In a network of multiple nodes, duty cycling mechanism depends on the harvested energy, consumed energy, distance between nodes and data aggregator/receiver, and the future harvested energy. In traditional \ac{iot}, nodes near the sink exhaust quickly due to the additional burden of relaying the data of far-away nodes in multi-hop communication. On the other hand, in \ac{ehiot}, nodes remain alive as long as they are receiving replenishable energy from the environment using energy harvesters.
Kansal et al.~\cite{kansal2004performance} present duty cycling mechanisms (for single node as well as multiple nodes) for sustainable performance in \ac{ehiot}. They describe a model for calculating the minimum size of the \ac{esu}/battery for sustainable operation of the embedded device. The duty cycle within a sensor node depends upon the average harvested as well as consumed energy in active and sleep modes.

The duty cycle can be adjusted such that the overall energy consumption does not exceed the overall harvested energy.
Similarly, if the harvested energy is increased, the duty cycle can also be raised to improve the performance within the given energy budget. Authors in~\cite{kansal2004performance} show the record of the battery voltage for more than two days which depicts that the node has adjusted its duty cycle in accordance with the harvested energy for sustainable operation. The authors also propose a duty cycling mechanism for a wireless network having multiple nodes. In a network configuration, determining sleep/awake time interval is challenging as the harvested energy for each node can be different due to its geographical location and orientation with respect to the energy source (e.g., solar energy). Therefore, the node attempting to transmit data will send a BEACON packet and wait for a certain time period for an ACK. It repeats the process until an ACK is received from the destination node. 

A mathematical model for duty cycling the sensor nodes according to the harvested energy is described in \cite{hsu2006adaptive}. It achieves the \ac{eno} and maximizes the system performance by adapting the dynamics of the replenishable energy source at run-time. It employs the \ac{ewma} algorithm to predict the harvested energy which is used to compute the duty cycle of the node.
Bouachir et al.~\cite{bouachir2017eamp} propose a \ac{mac} protocol for efficient energy utilization in cooperative \ac{wsns}. Their scheme uses the nodes' residual energy and data requirements to decide the active as well as sleep time periods of sensor nodes. As a result, it decreases the active time of nodes near the data aggregator and increases the active time of nodes away from data aggregator with the passage of time to balance the energy consumption of nodes. Therefore, it minimizes the problem of early depletion of nodes near the data aggregator, which reduces the coverage hole dilemma~\cite{khalifa2017coverage}.

Yang et al.~\cite{yang2015adaptive} propose a sensing scheduling scheme which dynamically adapts the sensing rate according to the available energy in the \ac{esu}.
They also propose a mathematical model for optimal sensing scheduling in energy harvesting sensors~\cite{yang2016optimal}. In contrast to the previous works that focus on the energy allocation to the sensors,~\cite{yang2016optimal} optimizes the sensing epoch depending on the energy budget. It presents the infinite and finite battery case and suggest an online scheduling policy that approaches the theoretical offline optimal scheduling mechanism. It also proposes a virtual energy harvesting sensing system to analyze the battery level which is helpful in deciding the sensing epoch.
An event-driven duty cycling mechanism is proposed in \cite{ali2016event} for power management of a road side monitoring unit. It harvests energy from the \ac{seh} and transmits data packets according to the events of the traffic flow on the road using an \ac{edf} algorithm. This implementation achieves lower energy consumption with longer lifetime of the sensing device.

\subsubsection{Duty cycling depending on the available and predicted harvested energy}
\begin{figure}[t!]
\includegraphics[width=8.5cm, height=6cm]{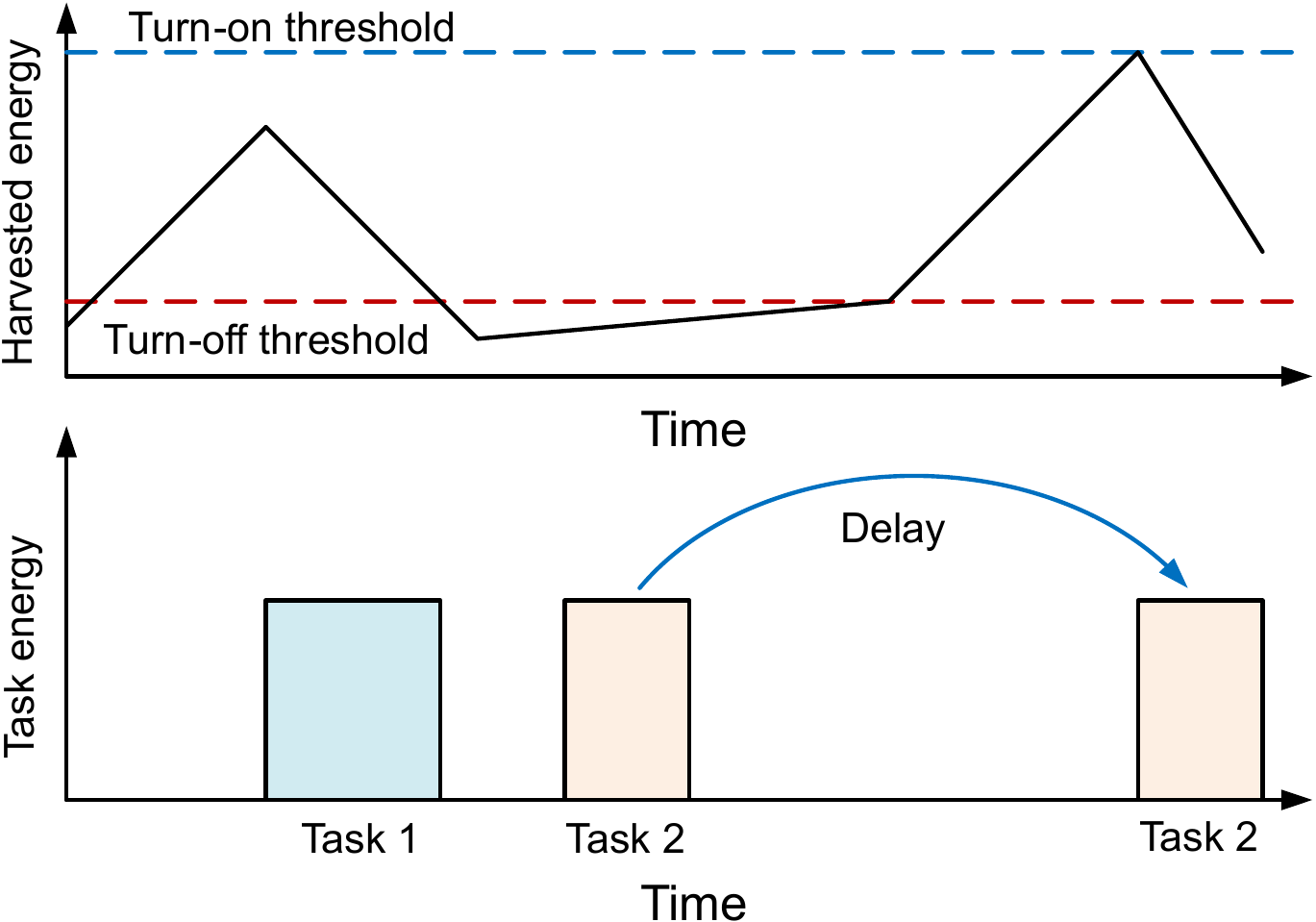}
\caption{Predicted harvested energy allows to delay the tasks execution during energy scarcity periods}
\label{fig:predicted_energy_motivation}
\end{figure}

In certain circumstances, the current harvested energy may not be sufficient to run an energy-intensive task on the node. Therefore, knowledge of the future incoming energy is important to delay the tasks until sufficient energy is available, without missing any deadline as illustrated in Fig.~\ref{fig:predicted_energy_motivation}. As shown in the figure, task 1 is executed as soon as it arrives at the node due to the higher energy availability than required for the execution of task. However, task 2 arrives when the available energy is lower than required for the execution of task. As there is a prediction of future harvested energy, the task is delayed until sufficient energy is available for its execution. In this way, the predicted harvested energy improves the energy utilization and minimizes the number of missed deadlines of tasks.
Moser et al.~\cite{moser2006lazy} present an algorithm for task scheduling in environmentally powered \ac{iot}. To the best of our knowledge, this is the first detailed work that addresses the task scheduling problem in \ac{ehiot}. In conventional \ac{iot}, the sensor nodes have a fixed energy supply (i.e. from a battery) and the only issue is to meet the deadlines of tasks for their execution on a single processor. In this case of abundant energy availability, various scheduling schemes such as \ac{asap} or \ac{alap} can be incorporated~\cite{moser2006lazy}. However, these schemes have certain drawbacks and can violate the deadline of tasks, when implemented on the energy harvesting node, which has intermittent energy availability. In the \ac{asap} algorithm, future incoming tasks can experience energy black-out periods (which results in deadline violations), whereas the \ac{alap} scheme can miss the deadlines, if the energy is limited near tasks deadlines. The \ac{edf} algorithm~\cite{andrews2000probabilistic} can solve this problem and schedules the tasks such that the task having an earlier deadline is executed first. However, \ac{edf} does not perform well in energy harvesting systems and violates deadlines due to energy consumption for less important tasks. This energy starvation results in the failure of execution of time-critical tasks with shorter deadlines. On the contrary, a scheme that hesitates in executing the tasks (until their deadlines) can perform well by conserving energy for time-critical tasks that have shorter deadlines~\cite{moser2006lazy}. 
The optimal start time of the task can be decided according the available energy and upcoming energy in the future.
Therefore, there is an opportunity to devise sophisticated algorithms for the prediction of future harvested energy. Then, this harvested energy profile determines the time slot for the execution of various tasks. Authors in~\cite{moser2006lazy} evaluate their technique using the energy from \ac{seh} and results show that it offers fewer deadline violations as compared to previous algorithms.

Authors in~\cite{moser2007real} extend the work of~\cite{moser2006lazy} with a detailed mathematical model and consider the practical considerations in implementing the algorithm on a real embedded device. They also propose an optimal start time for the execution of tasks depending upon their deadlines, energy consumption, stored energy and future harvested energy. 
However, an important factor in the calculation of optimal start time is the prediction of energy to be harvested in future.
In order to estimate the future harvested energy, authors in~\cite{moser2007real} propose an Energy Variability Characteristic Curve (EVCC) which gives an upper bound on the harvested energy in a certain time interval $\Delta$.
As a result, the harvested energy in any time interval $\Delta$ can easily be approximated, according to the given EVCC. However, another problem is to determine how much capacity the \ac{esu} should have, for perpetual operation of the sensor node in \ac{ehiot}. The answer lies in the maximum harvested energy, in addition to the consumed power in any particular time interval. This energy consumption can be calculated depending on the number of tasks arriving at the processor with common deadlines. 
Sommer et al.~\cite{sommer2019energy} propose a scheduling framework for various sensors (such as Global Positioning System (GPS), accelerometer, magnetometer, etc.) for perpetual tracking of flying foxes that travel long distances from their foraging camps in the search of food. The sensors are sampled based on the available and the future harvested energy. They also take into account the mobility and activity of flying foxes to trigger the next sensor sample. This technique ensures that maximum tracking accuracy is achieved within the given dynamic energy budget. The scheme in~\cite{sommer2019energy} is quite different from traditional tracking schemes that focus on minimizing the energy consumption with a fixed tracking accuracy. On the contrary,~\cite{sommer2019energy} emphasizes on achieving maximum tracking accuracy within the available energy budget which is replenished using on-board solar cells. The consideration of current and future energy levels while sampling the sensors ensures optimal time duration between successive samples, such that energy is neither depleted nor overflows (due to limited capacity of \ac{esu}) thanks to the future incoming energy from replenishable energy source (i.e., solar energy). Additionally, sensor sampling based on mobility ensures that enough energy is available to track the rest of the trip of flying foxes using the limited available energy.
Gy{\"o}rke et al.~\cite{gyorke2012application} exploit the knowledge about the environment to schedule non-equidistant samples, both is time as well as in space. The predicted harvested energy is taken into account to devise a conservative sampling approach when the future incoming energy is low. The proposed technique also uses the neighbour's information in deciding the duty cycle. It increases the duty cycle of nodes in vicinity of a place where an event has occurred. The other nodes operate at their usual duty cycle to conserve energy.

Akg{\"u}n et al.~\cite{akgun2013ehpbs} address the problem of duty cycling in cluster based \ac{ehiot} using \ac{tdma} slots. Traditional \ac{mac} protocols assign fixed \ac{tdma} slots to the cluster members for their data transmission. In contrast, the technique in~\cite{akgun2013ehpbs} uses the predicted harvested energy and available energy to assign \ac{tdma} slots to the nodes. In \ac{seh} based \ac{iot}, the node having maximum predicted harvested energy is assigned the next time slot in the day. On the other hand, at night, the node having maximum residual energy is given the next time slot.
Kooti et al.~\cite{kooti2012energy} present a task scheduling mechanism for \ac{seh} based \ac{iot} sensor nodes. Their scheme consists of two parts: offline scheduling and online scheduling. Firstly, the tasks are scheduled offline depending on the energy requirements and their criticality. Then, in the online scheduling phase, the tasks are adjusted according to the real-time available energy and future harvested energy, and are executed depending on their deadlines.
Renner et al.~\cite{renner2012adaptive} propose a scheduling algorithm on the basis of predicted harvested energy in \ac{iot} sensors. They present a harvested energy prediction model based on the changes in the previous harvested energy pattern in real-time, which reduces the prediction error and is applicable in real-world environments. Instead of periodic sampling, they exploit the periodic nature of the sun to adaptively trace the energy pattern and reduce the number of prediction updates.
Sommer et al.~\cite{sommer2013power} propose an accurate method for the estimation of \ac{soc} of the battery which can lead towards energy neutral scheduling of tasks. They use the current flow from the energy harvester and the battery voltage to estimate the current \ac{soc}. They employ a mathematical model called conflation~\cite{hill2011combine} to combine information from energy harvester as well as the battery to calculate the battery \ac{soc}.
Cui~\cite{cui2018solar} proposes the execution of tasks on the node using the predicted energy in \ac{seh} based \ac{iot}. He employs a recurrent long short term memory neural network to forecast the harvested energy. Later, this harvested energy is used to schedule tasks, which offers better performance than conservative and greedy approaches.

El Osta et al.~\cite{el2017real} propose to schedule periodic and aperiodic tasks separately, based on their priority. The periodic tasks are executed as soon they arrive at the scheduler if sufficient energy is available for their execution. On the other hand, aperiodic tasks are executed only if there is no periodic task in the queue and the \ac{esu} is not depleted. The aperiodic task consumes surplus energy that would have been wasted, if there is no other task ready for the execution. It is to ensure that the execution of aperiodic tasks does not affect the execution of future periodic tasks by consuming a significant amount of energy, which depletes the \ac{esu}.
Authors in~\cite{naderiparizi2015wispcam} use an \ac{rfid} reader to harvest energy for capturing and transmitting an image. The harvested energy is stored in an \ac{esu} to form fixed energy bursts to power the load (i.e., camera). It is more efficient to store the small amount of harvested energy instead of capturing large amount of energy as it may introduce delay in the execution of tasks. The \ac{esu} is also non-ideal, causing some leakage, which may reduce the usable energy budget. Therefore, it is more energy efficient to store the limited harvested energy to execute at least one atomic task as it takes a short build-up time and causes less leakage.
Instead of fixed energy bursts~\cite{naderiparizi2015wispcam}, Gomez et al.~\cite{gomez2016dynamic} introduce dynamic energy bursts that are matched with the requirements of the load. They introduce an \ac{emu} that arranges short energy bursts according to the requirements of the load and transforms small amounts of harvested energy into high-powered short energy bursts. Their \ac{emu} tracks the optimal power point of load and schedules the energy bursts accordingly. This results in efficient utilization of limited, variable and transient harvested energy to power the sensor nodes in \ac{ehiot}.

Zhang et al.~\cite{zhang2016solar} present a scheduling scheme for storage-less and converter-less energy harvesting systems in \ac{iot}. They formulate a system model for powering the sensor nodes using the \ac{seh} to minimize the deadline miss ratio. They develop a task scheduling algorithm using an artificial neutral network based on the harvested energy profile, the tasks priority and the energy budget. They also propose an integer nonlinear programming solution for optimal task scheduling on the node.
Severini et al.~\cite{severini2015energy} implement an energy-aware lazy scheduling algorithm on a low-cost and commercially available platform from Texas Instruments. This lazy scheduling algorithm delays the execution of tasks depending on their deadline and according to the available energy. Experiments are performed under varying incoming energy conditions. The results show the effectiveness of the scheduling algorithm and its suitability for real-time implementation in \ac{ehiot}.
Liu et al.~\cite{liu2015power} propose converter-less and storage-less operation of energy harvesting sensors. Due to the lower efficiency of the converter (40-90\%) and energy leakage of the \ac{esu}, the efficiency of the system is decreased. Authors in~\cite{liu2015power} design a hardware prototype using \ac{seh}, without converters and batteries, which performs better than conventional \ac{iot} systems. They also propose an intra-task scheduling algorithm which pre-empts the tasks according to their deadlines. The tasks are performed according to the available energy and their deadlines to meet the execution time requirement and use the energy efficiently. 
Zhang et al.~\cite{zhang2010energy} propose a scheduling mechanism for time-critical energy harvesting sensors. They present harvesting-aware speed selection of processor and radio transmission to manage the harvested energy and to meet the deadlines of tasks. They suggest two versions of their algorithm: central and decentralized, to be implemented according to the application environment. Simulations show that their scheme outperforms in terms of energy reserves than traditional schemes, which do not employ processor speed selection and transmission power control.

The available energy budget plays a pivotal role in scheduling the tasks on the nodes in \ac{ehiot}. Table~\ref{table_surveyed_papers} portrays that all of the previous task scheduling algorithms consider the available energy budget while deciding the execution of tasks to enhance the performance and minimizing deadline misses.

Fang et al.~\cite{fang2016optimal} propose a mathematical model for optimal scheduling in energy harvesting based mobile sensor nodes. As the harvested energy in mobile sensors varies with the change in energy source as well as with the mobility of embedded devices, it is challenging to devise an optimal scheduling algorithm. The sensor is switched to the sleep mode, when there is no task to be executed. After the arrival of new tasks, the corresponding sensor component is switched to active mode depending on the energy budget. Authors in~\cite{fang2016optimal} use a Lyapunov optimization framework to make control decisions that greedily minimize a bound on the drift-plus-penalty expression over fixed-length time slots. It has fair computational complexity and is easy to implement in real-time scenarios.
Pan et al.~\cite{pan2017lightweight} suggest a technique for the execution of tasks under unstable energy harvesting conditions. Their algorithm maximizes the progress of tasks using the information about the harvested energy. They also employ a routine which triggers sleep/wake-up events and start the execution of tasks automatically after resuming. 
Anagnostou et al.~\cite{anagnostou2018torpor} propose a power aware hardware scheduler for energy harvesting sensors. Instead of a traditional mechanism, where the harvested energy is stored in an \ac{esu}, they switch between the energy harvester directly and the \ac{esu}, depending on the harvested energy and task requirements. This mechanism leads to maximum utilization of the harvested energy even when the \ac{esu} is full to its capacity. Authors in~\cite{anagnostou2018torpor} implement an FPGA based prototype which shows that their algorithm achieves two times higher task execution rate under variable energy harvesting conditions, compared to the conventional schemes. 
Sanchez et al.~\cite{sanchez2017hybrid} propose a hybrid approach for energy management in \ac{ehiot}. This hybrid mechanism consists of battery states and the operating conditions of the nodes such as active, sleep and low duty cycle modes. In addition, they use decentralized control, which reduces the burden of controlling the whole network from a single node, extending the lifetime of the system.

The previous task scheduling algorithms employ conventional sensors, which consume energy during their operation. In addition, these sensors perform their operation using a reliable energy source (i.e., a battery) that provides continuous power for a longer duration. However, this scenario is entirely different for energy harvesting based sensing due to the dependence between energy and information, and intermittent execution of sensor node. Firstly, the energy harvester provides both energy as well as context information, and the signal needs to be sampled at suitable time instances according to the amount of context information and available energy. Secondly, the sensor node can run at most one atomic task at one time, due to the limited available energy stored in the capacitor. Therefore, it opens doors for dedicated task scheduling algorithms for energy harvesting based sensors. The objective of these task scheduling algorithms is to maximize the context detection accuracy as well ensure the \ac{eno} of sensor nodes in \ac{ehiot}.

Most of the previous task scheduling algorithms consider the predicted harvested energy, while scheduling the tasks on the node. This results in maximum utilization of the current and future harvested energy without missing the deadlines of tasks. In order to fully understand the functioning of the task scheduling algorithms, it is important to comprehend the energy prediction algorithms as well. Most of the harvested energy prediction algorithms use the previous harvested energy samples to estimate the future harvested energy profile. We discuss some of the previous energy prediction algorithms in the following subsection in detail.

\clearpage
\onecolumn
\begin{landscape}
\begin{longtable}[c]{|c|c|c|c|c|c|c|c|c|c|} 
  \caption{Summary of Task Scheduling Schemes in \ac{ehiot}}
  \label{table_surveyed_papers}\\
  \toprule
   & & \multicolumn{3}{c|}{\textbf{Scheduling policy}} & \multicolumn{2}{c|}{\textbf{Based on:}} & & & \textbf{Difficulty in using}\\
   \cmidrule{3-7}
  \textbf{Year and} & \textbf{Energy} & \textbf{DVFS} & \textbf{Decomp./} & \textbf{Duty} & \textbf{Predicted} & \textbf{Deadline} & \textbf{Performance} & \textbf{Performance} & \textbf{it in energy harvesting}\\
  \textbf{Reference} & \textbf{harvester} & & \textbf{Comb.} & \textbf{cycling} & \textbf{energy} & & \textbf{metric} & \textbf{assessment} & \textbf{based sensing}\\
  \midrule
  \endfirsthead 
  \toprule
     & & \multicolumn{3}{c|}{\textbf{Scheduling policy}} & \multicolumn{2}{c|}{\textbf{Based on:}} & & & \textbf{Difficulty in using} \\\cmidrule{3-7}
  \textbf{Year and} & \textbf{Energy} & \textbf{DVFS} & \textbf{Decomp./} & \textbf{Duty} & \textbf{Predicted} & \textbf{Deadline} & \textbf{Performance} & \textbf{Performance} & \textbf{it in energy harvesting}\\
    \textbf{Reference} & \textbf{harvester} & & \textbf{Comb.} & \textbf{cycling} & \textbf{energy} & & \textbf{metric} & \textbf{assessment} & \textbf{based sensing}\\
  \midrule
  \endhead 
2001 \cite{allavena2001scheduling} & General & \checkmark & & & & &  Remaining energy & Modeling & High \\
  \hline
2004 \cite{kansal2004performance} & Solar & & & \checkmark & & &  Latency & Simulation & High \\
 \hline
2006 \cite{hsu2006adaptive} & Solar & & & \checkmark & \checkmark (C) &  &  Energy utilization & Hardware & Medium \\
\hline
2006 \cite{moser2006lazy} & General & & & \checkmark & \checkmark (R) & \checkmark &  Deadline violation & Simulation & High \\
 \hline
2007 \cite{moser2007real} & Solar & & & \checkmark & \checkmark (C) & \checkmark &  Deadline violation & Simulation & High \\
 \hline
2008 \cite{liu2008energy} & Solar & \checkmark & & & & \checkmark  &  Deadline miss rate & Simulation & High \\
\hline
2009 \cite{liu2009adaptive} & Solar & \checkmark & & & & \checkmark &  Deadline miss rate & Simulation & High \\
\hline
2010 \cite{liu2010load} & Solar & \checkmark & & & & &  Deadline miss rate & Simulation & High \\
 \hline
2010 \cite{ravinagarajan2010dvfs} & Solar & \checkmark & & & & &  No. of executed tasks, Average & Hardware & High \\
 & & & & & & & accuracy, Energy consumption & & \\
 \hline
2010 \cite{zhang2010energy} & Water flow & & & \checkmark & & \checkmark &  Depleted nodes & Simulation & High \\
\hline
2011 \cite{ghor2011real} & Solar & & & \checkmark & & \checkmark  &  Deadline miss rate & Simulation & High \\
 & & & & & & & Capacity of energy storage & & \\
 \hline
2011 \cite{chetto2011real} & General & & & \checkmark & & &  Average idle time & Simulation & High \\
 & & & & & & & Feasible task set & & \\
 \hline
2011 \cite{audet2011scheduling} & Solar & & & \checkmark &  & \checkmark  & Battery charge level & Simulation & High \\
 & & & & & & & Energy violation rate & & \\
 \hline
 2012 \cite{zhu2012deos} & Solar & & \checkmark & & \checkmark (R) & \checkmark & Number of tasks executed, &  Hardware & Medium \\
 & & & & & & & missed deadlines & & \\
 \hline
 2012 \cite{liu2012harvesting} & Solar & \checkmark & & & \checkmark (C) & \checkmark &  Deadline miss rate & Simulation & High \\
& & & & & & & Minimum storage capacity & & \\
\hline
2012 \cite{yoo2012dynamic} & General & & & \checkmark & \checkmark (R) &  & End-to-end delay, BF, & Simulation & High \\
 & & & & & & & Packet delivery ratio & & \\
\hline
2012 \cite{gyorke2012application} & Solar & & & \checkmark & & & Energy consumption & Simulation & High \\
 \hline
2012 \cite{severini2012energy} & General & & & \checkmark & &  & Number of deadline violations & Simulation & High \\
 & & & & & & & First deadline violation & & \\
 \hline
2012 \cite{kooti2012energy} & Solar & & & \checkmark & \checkmark (R) & \checkmark & QoS violation count & Simulation & High \\
\hline
2012 \cite{renner2012adaptive} & Solar & & & \checkmark & \checkmark (C) & & Root mean square error & Simulation & High \\
\hline
2013 \cite{li2013dynamically} & Solar & & & \checkmark & & \checkmark  &  Time \& energy consumption & Simulation & High \\
\hline
2013 \cite{el2013nonclairvoyant} & General & & & \checkmark & & & Length of idle time &  Simulation & High \\
\hline
2013 \cite{akgun2013ehpbs} & Solar & & & \checkmark & \checkmark (C) & &  Network lifetime & Simulation & High \\
\hline
2013 \cite{lin2013framework} & Solar & \checkmark & &  & \checkmark (R) & &  Task drop rate & Simulation & High \\
\hline
2013 \cite{sommer2013power} & Solar & & & \checkmark & \checkmark (C) &  &  Battery SoC & Simulation & High \\
\hline 
2014 \cite{li2014task} & Solar & & & \checkmark & \checkmark (C) & &  No. of executed tasks & Simulation & High \\
\hline
2014 \cite{chetto2014optimal} & Solar & & & \checkmark & \checkmark (R) & \checkmark &  Time \& energy & Simulation & High \\
& & & & & & & constraint & & \\
\hline
2015 \cite{zhang2015deadline} & Solar & & & \checkmark & & \checkmark &  Deadline miss rate & Simulation & High \\
& & & & & & & Number of ready tasks & & \\
\hline
2015 \cite{naderiparizi2015wispcam} & RFID reader & & & \checkmark & &  & Power consumption & Hardware & Medium \\
& & & & & & & with time & & \\
\hline
2015 \cite{rao2015optimal} & Solar & & \checkmark & & \checkmark (R) & \checkmark & Number of tasks executed & Simulation & High \\
 \hline
2015 \cite{yang2015adaptive} & General & & & \checkmark & & & Mean Square Error & Simulation & High \\
\hline
2015 \cite{severini2015energy} & General & & & \checkmark & \checkmark (R) & \checkmark & Completed tasks & Hardware & Medium \\
\hline
2015 \cite{liu2015power} & Solar & & & \checkmark & & \checkmark & Deadline miss rate, & Hardware & Medium \\
 & & & & & & & Energy utilization & & \\
\hline
2016 \cite{tan2016dynamic} & General & \checkmark & &  & & \checkmark & Average overhead & Simulation & High \\
 & & & & & & & Average busy period & & \\
\hline
2016 \cite{housseyni2016real} & Solar & & & \checkmark & \checkmark (C) & \checkmark &  Deadline success ratio & Simulation & High \\
\hline
2016 \cite{gomez2016dynamic} & Solar & & & \checkmark & & & System efficiency & Hardware & Medium \\
\hline
2016 \cite{yang2016task} & General & & & \checkmark & & \checkmark &  Deadline miss rate, & Simulation & High \\
 & & & & & & & energy violation rate & Hardware & \\
 \hline
2016 \cite{maeda2016dynamic} & Solar & & & \checkmark & & & No. of packets sent to BS, & Simulation & High \\
& & & & & & & Delay & & \\
\hline
2016 \cite{zhang2016solar} & Solar & & & \checkmark & \checkmark (C) & \checkmark & Deadline miss ratio, & Hardware & Medium \\
& & & & & & & Energy utilization efficiency & & \\
\hline
2016 \cite{fang2016optimal} & General & & & \checkmark & & & Data queue size, & Simulation & High \\
 & & & & & & & Energy queue size & & \\
\hline
2016 \cite{yang2016optimal} & General & & & \checkmark & & & Sensing performance, Infeasible & Simulation & High \\
 & & & & & & & sensing epoch, Battery overflow & & \\
\hline
2016 \cite{ali2016event} & Solar & & & \checkmark & & \checkmark & Power consumption, & Hardware & Medium \\
 & & & & & & & Battery lifetime & & \\
\hline
2017 \cite{el2017real} & General & & & \checkmark & & \checkmark & Deadline miss rate, & Simulation & High \\
 & & & & & & & response time & & \\
 \hline
2017 \cite{bouachir2017eamp} & General & & & \checkmark & & \checkmark & Harvested to consumed & Simulation & High \\
& & & & & & & energy rate & & \\
\hline
2017 \cite{pan2017lightweight} & General & & & \checkmark & & & Energy and time overhead & Simulation & High \\
& & & & & & & Energy consumption & & \\
\hline
2017 \cite{sanchez2017hybrid} & General & & & \checkmark & & & Remaining energy of node & Hardware & Medium \\
\hline
2018 \cite{huang2018adaptive} & Solar & & & \checkmark & & & Expected reward & Simulation & High \\
\hline
2018 \cite{anagnostou2018torpor} & Solar & & & \checkmark & & \checkmark & Energy efficiency, & Hardware & Medium \\
& & & & & & & Execution rate & & \\
\hline
2018 \cite{cui2018solar} & Solar & & & \checkmark & \checkmark (C) & & Task completion rate & Simulation & High \\
\hline
2019 \cite{sommer2019energy} & Solar & & & \checkmark & \checkmark (C) & \checkmark & Average tracking error & Simulation & High \\
  \bottomrule
\end{longtable}
\end{landscape}
\clearpage
\twocolumn

\subsection{Algorithms for energy prediction in \ac{ehiot}}
\label{subsec:predicting_harvested_energy_SEH_IoT}
In order to ensure \ac{eno}, the sensor nodes need to consume the harvested energy in such a way that the current node operation is not affected, and the future tasks do not run out of energy. Therefore, information about the future harvested energy is important to schedule the energy consumption proactively for sustainable operation of the system. In the literature, there are various harvested energy prediction models that utilise previous energy samples, weather conditions and seasonal trends to compute the future harvested energy in energy harvesting based \ac{iot}. Table~\ref{table_predicted_energy} comprehensively presents some of the harvested energy prediction algorithms that employ statistical, probabilistic and machine learning models to predict the future harvested energy.
On a normal sunny day, the harvested energy from a solar powered node is highest at noon, and decreases at dawn and dusk, finally reaching zero at night~\cite{piorno2009prediction}, due to the non-availability of sunlight. Knowing this overall pattern of \ac{seh}, the future solar harvested energy can be predicted using the previous energy pattern. Kansal et al.~\cite{kansal2007power} present a harvested energy prediction model based on \ac{ewma} for solar powered \ac{iot}. Their model relies on the intuition that the harvested energy in a particular day at a given time slot is similar to that of the energy harvested in the previous days at the corresponding time slots. Therefore, the harvested energy in a particular time slot is calculated by accumulating the weighted average of harvested energy in the previous days in the same time.
\ac{ewma} algorithm awards higher weight to the recent energy values and exponentially decreases weight for the previous energy samples to calculate the future harvested energy in solar powered \ac{iot}. The weight is calculated dynamically using the real previous energy traces, which provide the lowest value of error. However,~\cite{kansal2007power} gives significant prediction error when there is a sudden change in the weather. It is due to the reason that the \ac{ewma} scheme does not take the seasonal weather trends and diurnal cycles into account. Hassan et al.~\cite{hassan2012solar} propose an energy prediction model for solar powered \ac{iot}, which takes into account the sudden changes in the environment. It also takes into account the seasonal and diurnal cycles of the solar energy. 
However, this scheme is more computationally complex as it takes multiple parameters into account and costs more energy as well as processing time, incurring delay in the system. Another technique which considers weather conditions for predicting the harvested energy in solar powered \ac{iot} is presented in~\cite{sharma2010cloudy}. It presents a model to predict solar as well as wind harvested energy. The method takes the data from the weather forecast stations to predict the energy to be harvested in future time slots. However, this scheme is also computationally complex and depends on another source, which increases the cost of the system. Additionally, receiving weather data and processing it on an energy-constrained miniaturized sensor node hinders the execution of other time-critical tasks. 
Piorno et al.~\cite{piorno2009prediction} present a prediction algorithm for \ac{seh} which depends on \ac{ewma} and takes into account the sudden and abrupt changes in weather conditions. They propose to use the weighting factor depending on the solar conditions of the current day relative to the previous days.
Cui~\cite{cui2018solar} proposes a \ac{seh} prediction algorithm for sensor nodes in \ac{ehiot}. It employs a recurrent long short term memory neural network to forecast the harvested energy. However, this is a complex method that has higher cost in terms of energy, time, memory requirement and computational resources.

\begin{table}[t!]
\caption{Previous algorithms in the literature for predicting the harvested energy in solar powered \ac{iot} sensor node}
\label{table_predicted_energy}
\centering
\begin{tabular}{llll}
\toprule
\textbf{Year} & \textbf{Reference} & \textbf{Input parameters} & \textbf{Method}\\
\midrule
\midrule
2007 & \cite{kansal2007power} & Previous samples & \ac{ewma} \\
\hline
2008 & \cite{moser2008robust} & Previous minimum energy & Worst-case energy \\
 & & & prediction\\
\hline
2009 & \cite{piorno2009prediction} & Weather conditions from & Weather-conditioned \\
& & recent past samples & moving average \\
\hline
2010 & \cite{sharma2010cloudy} & Weather forecast & Quadratic solar \\
 & & & power model \\
\hline
2011 & \cite{ventura2011markov} & Multiple energy harvesters & Markov model \\
\hline
2012 & \cite{cammarano2012pro} & Previous energy profiles & Profile energy \\ 
 & & & prediction \\
\hline
2016 & \cite{cammarano2016online} & Previous energy profiles & Profile energy\\
& & & prediction\\
\hline
2016 & \cite{kosunalp2016new} & Past observations, & Q-learning \\
 & & Current weather & \\
\hline
2016 & \cite{zou2016energy} & Previous samples & Curve fitting \\
\hline
2019 & \cite{geissdoerfer2019getting} & Global horizontal & Austronomical \\
  &   &  irradiance &  model \\
\bottomrule
\end{tabular}
\end{table}

Most of the previous harvested energy prediction schemes~\cite{kansal2007power,hassan2012solar,sharma2010cloudy}, forecast the future solar harvested energy for the next single time slot. However, occasionally, it is also important to estimate the harvested energy for future $N$ (where $N>1$) time slots. Moser et al.~\cite{moser2008robust} present a harvested energy prediction algorithm that calculates the harvested energy for future $N$ time slots. Their algorithm considers the harvested energy in previous time slots of length $N$. Then, the worst case harvested energy in the previous time slots is considered as the predicted energy for the future slots. However,~\cite{moser2008robust} does not take the fast changing weather into account which results in higher prediction errors in swift weather changing environments. It is also a pessimistic approach, as it considers the worst case harvested energy only. Cammarano et al.~\cite{cammarano2012pro,cammarano2016online} propose a prediction model using harvested energy profiles of previous days. They store different types of energy profiles (like sunny, partially sunny, cloudy, etc.) and compare the initial values of currently harvested energy with the stored energy profiles. The stored energy profile having highest correlation with the current harvested energy is considered as the predicted energy profile for the rest of the day. However, this scheme also burdens the miniaturized \ac{iot} nodes with complex computation and data storage requirements.
Ventura et al.~\cite{ventura2011markov} present an energy harvesting and consumption algorithm for body sensor networks using a Markov model. Their algorithm considers multiple types of energy harvesters and predicts the future states of nodes in terms of energy level depending on the probabilistic model based on previous energy samples. Authors in~\cite{kosunalp2016new} propose a Q-learning based solar harvested energy prediction model using previous energy samples and current weather conditions. This results in lower prediction error than conventional \ac{ewma}. As the harvested energy from \ac{seh} depends on luminous intensity, Zou et al.~\cite{zou2016energy} propose to predict the locations of sunlight for next time interval to predict the harvested energy. They use a piecewise least squares curve fitting estimation using the previous samples to estimate the future harvested energy. Kai et al.~\cite{geissdoerfer2019getting} predict the future solar harvested energy using global horizontal irradiance and solar cell's characteristics. In order to compensate for deviations from the actual values, their scheme compares the predicted harvested energy values with the previous harvested energy. 

In summary, there are various energy prediction algorithms for solar powered \ac{iot}, which take into account the previous harvested energy values, weather forecasts and previous energy profiles, to correctly predict the future harvested energy, as shown in Table~\ref{table_predicted_energy}. However, these algorithms, when implemented on the node, consume a significant amount of harvested energy during their execution. Therefore, in addition to prediction accuracy, the cost in terms of energy and computational complexity must also be explored. Such an analysis will provide the real picture about the models and will identify the algorithm which provides best energy prediction results, while executing within the limited energy and computational resources.

\subsection{Summary and insights}
We comprehensively survey and analyse previous task scheduling algorithms in \ac{ehiot} to enhance the lifetime of sensor nodes. However, none of the previous algorithms employ energy harvesting based sensing; instead they exploit conventional sensors for monitoring the desired physical parameter. It results in significant energy consumption compared to energy harvesting based sensors as discussed in Section~\ref{sensing_and_energy_harvesting}. Therefore, keeping in view this new class of sensors (i.e., energy harvesting based sensing), the previous task scheduling algorithms need to be revised or new dedicated task scheduling algorithms should be proposed to allow the sustainable operation of sensor nodes in \ac{ehiot}. We critically analyse the previous task scheduling schemes and explore their applicability for energy harvesting based sensors in the following section. In addition, we rigorously discuss the opportunities for transforming the conventional task scheduling algorithms for the emerging class of energy harvesting based sensors.

\begin{figure}[t!]
\centering
\includegraphics[width=8cm, height=8cm]{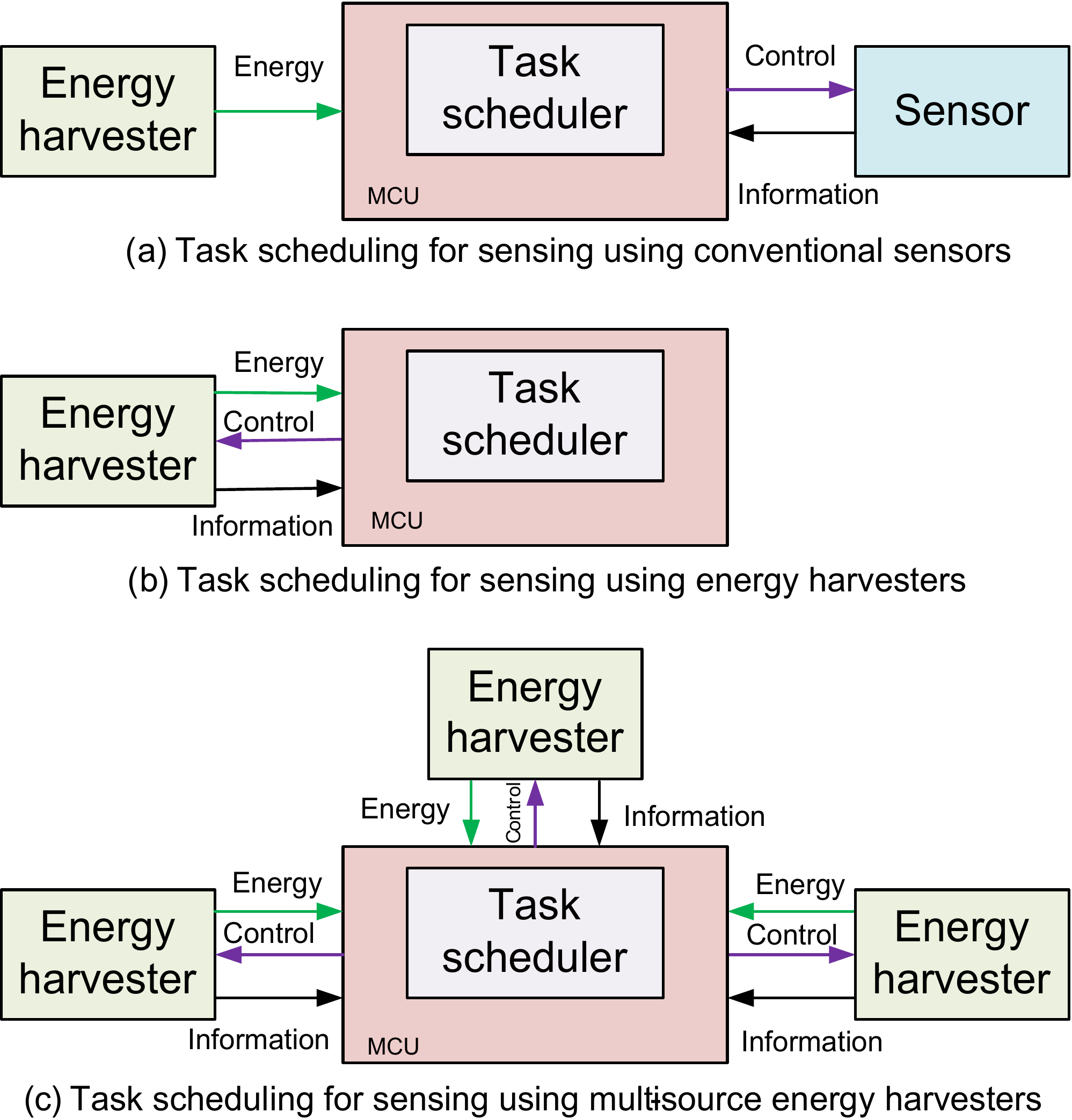}
\caption{Interaction of task scheduler with (a) conventional sensor, (b) energy harvesting based sensor, and (c) multi-source energy harvesting based sensors}
\label{fig:task_scheduling_future}
\end{figure}

\section{Critical Analysis of Task Scheduling Algorithms for Energy Harvesting based Sensing}
\label{sec:challenges_and_opputunities}
We discussed that task scheduling algorithms manage the execution of tasks on the node to extend the lifetime of sensor nodes in conventional \ac{ehiot}. However, considering the new category of sensors (i.e., energy harvesting based sensing), we explore the applicability of task scheduling algorithms in this new paradigm. There is a fundamental difference between conventional and energy harvesting based activity sensors in terms of their operation and energy consumption. Conventional sensors consume energy while in operation (see Fig.~\ref{fig:task_scheduling_future}(a)), whereas, energy harvesting based sensors generate energy, in addition to providing a signal that can be used for detecting the underlying activities, as depicted in Fig.~\ref{fig:task_scheduling_future}(b). As energy and context information from the energy harvester are strongly correlated, aggressive sampling of the harvesting signal may not be required, which opens new directions for devising task scheduling mechanism compared to conventional activity sensors. In addition, the harvested energy can act as a trigger signal to decide the next sensing epoch depending on the variation in the harvesting signal. Furthermore, the variation in the harvesting signal can be translated as the change in ongoing activity, using few initial samples of the energy harvesting signal. The task scheduling becomes more challenging in multi-source energy harvesting based sensing, as shown in Fig.~\ref{fig:task_scheduling_future}(c), due to the different amount of energy and context information from each harvester unit. Depending upon the type of application, one energy harvester can provide higher context information compared to others, in certain physical conditions, which needs to be tracked in real-time. This results in higher activity detection accuracy without consuming significant energy, for example, to acquire the signal at a substantially high sampling rate or to sample all energy harvesting signals. In addition, the amount of harvested energy from each energy harvester is unique depending upon the environmental conditions and properties of the transducer. This energy can be combined using DC-DC converters or individual harvesters can be selected according to the amount of generated energy, to avoid the energy conversion losses~\cite{wang2015storage}.

\subsection{Challenges of current task scheduling techniques}
Most of the existing task scheduling algorithms are complex and require significant amount of energy during their operation on the miniaturized and resource-constrained sensor node. Decomposing and combining the tasks is not applicable for batteryless energy harvesting based sensors due to their limited energy budget that can, at most, run one atomic tasks at a time. Secondly, \ac{dvfs} algorithms need complex hardware circuits to provide multiple voltage levels to individual components of the sensor node. It may require multiple \ac{esu}s in batteryless sensor nodes which increase the energy losses, cost and form factor, and decrease the usable energy. Most of the previous task scheduling algorithms focus on enhancing the lifetime of the system, ignoring the performance in terms of activity detection accuracy. These scheduling algorithms do not take into account the correlation between energy and context information, which may help in deceasing the sampling rate of the harvesting signal, resulting in less energy consumption, in contrast to conventional activity sensors. In addition, current mechanisms rely on sequential functioning of the program, which is not applicable for intermittent operation in batteryless sensors, which experience frequent energy blackouts. Timing failure is another issue in batteryless devices due to frequent power failures and energy scarcity periods, in contrast to conventional battery operated sensing devices. Furthermore, devising task scheduling algorithms for multi-source energy harvesters is more challenging due to the varying amount of harvested energy and context information from each harvester unit in real-time. Moreover, the predicted harvested energy plays an important role in devising task scheduling algorithms, as discussed in Section~\ref{Scheduling_in_ehiot}. However, there is no energy prediction algorithm for kinetic, thermal and RF energy, in particular, for applications which involve mobility. Furthermore, there is no existing mechanism to manage the additional harvested energy in energy positive sensors, which may lead towards \ac{eno} of \ac{ehiot} sensors. Finally, most of the current task scheduling algorithms are validated through computer simulations without implementing on real hardware test bed.

Based on the comprehensive discussion of existing task scheduling algorithms in Section~\ref{Scheduling_in_ehiot} and aforementioned challenges, we describe the possible solutions and future directions for incorporating energy harvesting based sensors in the next subsection.

\subsection{Possible solutions}
Task scheduling algorithms should be devised to focus both on enhancing the context information as well as the system lifetime. Among the available options, duty cycling is the most appropriate choice for scheduling the tasks on the miniaturized and resource-constrained sensor node due to its inherent lower complexity and ease of implementation. It is also compatible with the batteryless sensors due to their transient and intermittent operation. In addition, it is also important to devise harvested energy prediction algorithms to allow scheduling of tasks according to the harvested energy. As energy prediction algorithms consume a significant amount of energy during their operation, another approach is to devise task scheduling algorithms without quantifying the future energy, which can reduce the energy consumption of miniaturized sensor nodes, resulting in sustainable operation of the system. It is also important to analyse the complexity and overhead in terms of energy for running task scheduling algorithms before implementation on real resource-constrained sensor nodes. Depending upon the type of application and nature of the environment in which energy harvesters are employed, task scheduling algorithms can be adapted in real-time. Furthermore, in case of multi-source energy harvesters, it is important to switch between the most suitable energy harvesters depending upon the context information and harvested energy from the individual transducer output signals in real-time. In energy harvesting based sensing, context information and the amount of harvested energy are correlated. Therefore, energy can be saved by reducing the sampling rate of the sensing signal during the steady state, without losing context detection accuracy. Moreover, the additional harvested energy can be employed to run other pending tasks on the node, such as data processing and communication, which leads towards the potential of \ac{eno} in \ac{ehiot}. Therefore, new operating systems and programming frameworks are needed for running energy harvesting based sensors, which take into account the intermittent operation of batteryless sensors, to resolve the checkpointing and timing failure issues, which are not present in conventional battery-based devices. Finally, there is a potential of task scheduling algorithms for energy harvesting based sensors, to meet the objectives of both maximized sensing performance as well as ensuring \ac{eno} of sensor nodes in \ac{ehiot}.

\section{Future Research Directions}
\label{Future_Research_Directions}
We discussed that energy harvesters can be employed as a simultaneous source of energy as well as context information. Since this new class of sensors has inherent differences than conventional sensors, it brings opportunities for dedicated energy management algorithms for its perpetual operation. Therefore, conventional task scheduling algorithms need to be revised due to the correlation between energy and information in energy harvesting based sensing.
As the harvested energy is limited, task scheduling algorithms are required to execute tasks on the node according to harvested energy profile, for achieving \ac{eno} of sensor nodes in \ac{ehiot}. There are various challenges in devising efficient task scheduling algorithms for energy harvesting based sensing due to the varying and intermittent harvested energy that puts further constraints in case of batteryless sensors. The goal of task scheduling algorithms is to enhance the activity detection performance as well as the operational lifetime of sensor nodes in \ac{ehiot}. Based on the previous discussion, future research directions for ensuring higher activity detection accuracy as well as achieving \ac{eno} of sensor nodes are described in the following subsections in detail.

\subsection{Exploring optimal sampling frequency of sensing signals}
In order to extract context information, the sensing signals from the energy harvesting circuit are sampled, whereas processing can be done offline or on the device, depending on the energy budget. The sampling frequency of the energy harvesting signal is an important parameter which plays an important role in the activity detection accuracy. Increasing the sampling frequency results in higher energy consumption with the advantage of higher activity detection accuracy. Therefore, there is a potential to study the optimal sampling frequency of various sensing signals in terms of energy consumption as well as activity detection accuracy. As the harvested energy varies according to the type of application, the sampling frequency also varies accordingly which needs to be studied in detail.

\subsection{Maximum Power Point (MPP) tracking}
As the output voltage and current of the energy harvesting transducer changes due to the environmental conditions, its \ac{mpp} also varies swiftly under the external stimuli. \ac{mpp} tracking of \ac{keh} transducer is more challenging than solar cells, due to rapid output voltage/current fluctuations in the former compared to the relatively slow variations in the latter. Therefore, sophisticated hardware modules are required which can dynamically track its \ac{mpp} at a high frequency in run-time~\cite{sandhu2020optimal}. However, it will also consume more energy in sampling the signal at a higher frequency to find its \ac{mpp} voltage. Therefore, there is a potential to study the optimal \ac{mpp} tracking frequency, amount of energy consumed in tracking and additional harvested energy, to estimate the overall gain.

\subsection{Employing multi-source energy harvesters as a simultaneous source of energy as well as context information}
In order to ensure the sustainable operation of sensor nodes in \ac{ehiot}, the harvested energy should be sufficient to power the sensor's hardware without the need of any external energy source (i.e., a battery). 
Occasionally, the harvested energy from a single energy harvester is not sufficient to continuously power each module of the sensor node. For example, \ac{seh} can not provide sufficient energy during night and darkness. On the other hand, the harvested energy from \ac{keh} is very small during lower vibrations/movements, such as sitting and standing in human activity recognition applications. Therefore, multi-source energy harvesters (e.g., \ac{seh}, \ac{keh}, \ac{teh} and \ac{rfeh}) can be employed to harvest higher energy as well as extract rich activity information. The energy from these harvesters can be accumulated to power a sensor node to achieve \ac{eno}. In addition, the signals from multi-source energy harvesters can be fused to extract rich context information. 
For example, while \ac{keh} provides information about the movement/activity, \ac{seh} can be used to identify the indoor and outdoor environments, depending upon the amount of harvested energy, which can be employed in the localization applications. However, if a signal does not possess information about the underlying activity, its fusion with other signals may increase the cost, complexity and energy consumption of the system.
In summary, there is a potential to study the amount of harvested energy and context detection accuracy from multi-source energy harvesters in various applications, such as human activity recognition, gait recognition and transport mode detection.

\subsection{Scheduling framework for energy harvesting based sensing}
In order to ensure \ac{eno} of sensor nodes and achieve higher activity detection accuracy, the harvested energy should be consumed efficiently to run a maximum number of tasks on the sensor node in \ac{ehiot}. In energy harvesting based sensors, the execution of tasks is a function of context information, harvested energy and required energy.
However, the previous scheduling schemes employ conventional motion sensors, instead of using the energy harvesters as activity sensor and energy source simultaneously. Furthermore, multi-source energy harvesters can be used to extract higher energy as well as rich context information, which brings new challenges due to the different amount of energy and context information from each harvester unit. This new class of sensors (i.e., sensing using energy harvesters) brings opportunities for revised task scheduling schemes, due to the correlation between harvested energy and context information. Eventually, it will enable the perpetual operation of sensor nodes along with higher detection accuracy of the underlying activity, without the need of any external depletable energy source (i.e., a battery). Considering energy harvesters as activity sensors and source of energy simultaneously, the key research challenges for devising task scheduling algorithms are as follows:
\begin{itemize}
    \item Devising a duty cycling algorithm for the sensor node according to the harvested energy profile, which provides highest activity detection accuracy with minimum power consumption.
    \item Studying the relation between the harvested energy and the sampling frequency of the energy harvesting signal.
    \item In order to acquire actual and real pattern of the energy harvesting signal, the energy harvesting transducer can be disconnected from the capacitor/load during sampling the signal for context detection applications. However, it may result in lower harvested energy with the advantage of higher activity detection accuracy, which needs to be studied in detail.
    \item In case of multi-source energy harvesters, there is a need to devise an optimal sampling frequency for each energy harvesting signal.
    \item Exploring the information gain of each energy harvesting signal in multi-source energy harvesters and fusing the most information-rich signals, that provide highest activity detection performance, with minimum energy consumption.
    \item Processing the energy harvesting signals and extracting the dominant features on the resource-constrained sensor node. This may result in lower energy consumption in data transmission (due to smaller feature set) with higher energy consumption in processing the data on the \ac{iot} sensor node.
    \item Implementing the classification algorithm on the sensor node for online (real-time) activity detection, resulting in autonomous operation of sensor nodes in \ac{iot}.
    \item Devising a scheduling framework in batteryless sensors that work intermittently under limited and un-reliable harvested energy.
    \item Studying the trade-off between the online and offline processing of various sensing signals, collected from the energy harvesting transducers.
    \item In case of lower harvested energy, the collected data or extracted features may not be transmitted to the server for offline processing. Therefore, array of energy harvesters can be employed to harvest higher energy with the overhead of increased cost and form factor.
\end{itemize}
In summary, considering the aforementioned research challenges while devising task scheduling schemes will ensure the ultimate goals of higher context detection accuracy as well as \ac{eno} of sensor nodes in \ac{ehiot}.
\begin{table}[t!]
\caption{Comparison between harvested energy in \ac{seh} and \ac{keh} in the context of energy prediction}
\label{table_Comp_Pred_EH}
\centering
\begin{tabular}{ll}
\toprule
\textbf{Solar energy harvesting} & \textbf{Kinetic energy harvesting}\\
\midrule
\midrule
Periodic & Aperiodic\\
\hline
Relatively stable & Relatively unstable\\
\hline
Easily predictable & Difficult to predict\\
\hline
Higher energy than \ac{keh} & Lower energy than \ac{seh}\\
\hline
Easy to design the power & Difficult to design the power\\
conditioning circuit & conditioning circuit\\
\hline
Power is generated in the & Power is generated in the \\
presence of light & presence of vibrations\\
\hline
Relatively less noise in the  & Relatively more noise in the\\
signal & signal \\
\hline
Generates DC voltage & Generates AC voltage\\
\bottomrule
\end{tabular}
\end{table}
\subsection{Predicting the harvested energy in \ac{ehiot}}
In contrast to \ac{seh}, the harvested energy in \ac{keh} is highly fluctuating, which varies quickly according to the underlying stress/vibrations. This unstable output voltage poses more challenges to devise a prediction model for the harvested energy. There is a significant difference between the energy pattern in \ac{seh} and \ac{keh}, as listed in Table~\ref{table_Comp_Pred_EH}. The harvested energy in \ac{seh} is predictable due to its overall known pattern as described in Section~\ref{subsec:predicting_harvested_energy_SEH_IoT}. However, in contrast to \ac{seh}, the harvested energy in \ac{keh} can not be predicted by merely accumulating the previous weighted energy samples, as in \ac{ewma}~\cite{kansal2007power}, especially for long term energy predictions. The reason behind it lies in the sudden changes in the harvested energy due to the change in the nature of the underlying vibration source. The harvested energy pattern in \ac{keh} based \ac{iot} depends upon the type of application which demands dedicated models for harvested energy prediction for each use case. If the target application is human activity recognition, the \ac{keh} energy is non-identical in different types of activities~\cite{khalifa2018harke}, including walking, standing, running, going upstairs/downstairs, etc. Similarly, if the target application is transport mode detection~\cite{lan2018entrans}, the harvested energy depends on the mode of transportation. Therefore, \ac{keh} demands different energy prediction models due to the distinct harvested energy pattern in each application. Furthermore, in contrast to static \ac{seh} applications (except~\cite{sommer2019energy}), \ac{keh} is typically deployed in a mobility scenario, and has various states in most of the context detection applications, including human activity recognition and transport mode detection. Therefore, in order to predict the harvested energy in \ac{keh}, the mobility of the energy harvester must be taken into account to achieve higher prediction accuracy. Similarly, dedicated energy prediction algorithms are required for other energy harvesters, such as \ac{teh} and \ac{rfeh} to estimate the harvested energy for execution of the tasks on sensor nodes efficiently.

The predicted harvested energy plays an important role in scheduling the tasks in \ac{ehiot} sensor nodes. As the capacity of \ac{esu} is limited (due to small-sized capacitors/batteries), the harvested energy can overflow if the stored energy is not consumed in executing the tasks beforehand, which results in the wastage of resources. For example, when the battery is fully charged, there is no room to store the incoming harvested energy, which results in energy wastage, if it is not properly utilised. The solution is to utilise the maximum energy, when the \ac{esu} is charged to its capacity as well as there is a prediction of future harvested energy. It results in the efficient employment of resources and the maximum utilization of energy in executing the tasks within their deadlines. Similarly, if the \ac{esu} is depleted, the tasks can be delayed, according to the predicted harvested energy, to achieve a minimum performance level.

\subsection{Batteryless \ac{iot}}
Conventional \ac{ehiot} employ batteries to store the harvested energy and power the sensor nodes. However, batteries are costly, bulky, toxic and have a limited lifetime~\cite{hester2017future}. A promising solution is to use capacitors to store the harvested energy. As capacitors generally have higher leakage and lower energy storage capacity compared to batteries, task scheduling with a capacitor-based \ac{esu} is more challenging. Intermittently charged capacitors restrict the continuous utilization of energy, which further restrain the frequent execution of tasks. Therefore, there is a potential of task scheduling algorithms that take into account the intermittent operation of load/sensor node to enable the autonomous operation of batteryless sensor nodes in \ac{ehiot} with higher context detection performance.

\section{Conclusion}
\label{conclusion}
Energy harvesters are employed to power sensor nodes in \ac{ehiot} to replace the conventional manually rechargeable batteries that hinder their widespread adaptability and pervasive deployment. In addition to energy generation, recently energy harvesters have been used as sensors for context detection. 
This saves significant energy that would otherwise be used for powering conventional activity sensors. 
Using the energy harvester as a simultaneous source of energy and information enables \textit{energy positive sensing}, which harvest higher energy than required for signal acquisition for context detection.
However, the harvested energy is still not sufficient to allow the \ac{eno} of sensor nodes in \ac{iot}. In order to ensure the sustainable operation of sensor nodes, the precious harvested energy needs to be consumed very efficiently for running the operational tasks on the nodes.
In this survey paper, we comprehensively analyse the previous task scheduling based energy management algorithms for \ac{ehiot} sensors. We critically analyse the challenges in incorporating the emerging class of energy harvesting based sensors in the conventional task scheduling algorithms. Based on the extensive study of the literature, we rigorously review the need for revised task scheduling algorithms for energy positive sensors and provide potential solutions. Finally, we present future research directions towards the goal of enabling the sustainable and autonomous operation of batteryless sensor nodes in \ac{ehiot}.

\ifCLASSOPTIONcaptionsoff
  \newpage
\fi
\bibliographystyle{IEEEtran}
\bibliography{main_v11.bbl}
\end{document}